\documentclass[fleqn,usenatbib]{mnras}
\include{notations}
\usepackage{slashed}
\usepackage{graphicx}
\usepackage{grffile}
\usepackage[normalem]{ulem}
\usepackage{dcolumn}
\usepackage{xcolor}
\usepackage{booktabs}
\usepackage{amsmath}
\usepackage{amssymb}
\usepackage{slashed}
\usepackage{todonotes}
\usepackage{comment}
\usepackage{subfigure}
\usepackage{lastpage}
\usepackage{multirow}

\newcolumntype{M}{R@{${}\pm{}$}L}

\usepackage{multirow}
\usepackage{arydshln}
\usepackage[T1]{fontenc}
\usepackage{ae,aecompl}
\usepackage{lineno}
\usepackage{orcidlink}

\usepackage{txfonts}

\newcommand{\ahod}{\textsc{AbacusHOD}}

\newcommand\vR{\boldsymbol{r}_p}
\newcommand\vr{\boldsymbol{r}}

\newcommand\vk{\boldsymbol{k}}

\newcommand\intk{\int \frac{d^3\vk}{(2\pi)^3}}
\newcommand\intkp{\int \frac{d^3\vk'}{(2\pi)^3}}
\newcommand\intvkperp{\int \frac{d^2\vk_\perp}{(2\pi)^2}}
\newcommand\intvkpperp{\int \frac{d^2\vk_\perp'}{(2\pi)^2}}
\newcommand\intkperp{\int \frac{dk_\perp \, k_\perp}{2\pi}}
\newcommand\intkpar{\int_{-\infty}^\infty \frac{dk_\parallel}{2\pi}}

\title[GGL+GC systematics]{Redshift evolution and covariances for joint lensing and clustering studies with DESI Y1}

\date{\today}

\pubyear{2022}

\author[Yuan et al.]{\parbox[t]{0.99\textwidth}{\vspace{-0.7cm}
Sihan Yuan,$^{1,2}$\thanks{E-mail: sihany@stanford.edu}\orcidlink{0000-0002-5992-7586}
Chris Blake,$^3$\thanks{E-mail: cblake@swin.edu.au}
Alex Krolewski,$^{4,5}$
Johannes Lange,$^6$\orcidlink{0000-0002-2450-1366}
Jack Elvin-Poole,$^7$
Alexie Leauthaud,$^8$\orcidlink{0000-0002-3677-3617}
Joseph DeRose,$^9$\orcidlink{0000-0002-0728-0960}
Jessica Nicole Aguilar,$^9$
Steven Ahlen,$^{10}$\orcidlink{0000-0001-6098-7247}
Gillian Beltz-Mohrmann,$^{11}$
David Brooks,$^{12}$
Todd Claybaugh,$^{9}$
Axel de la Macorra,$^{13}$\orcidlink{0000-0002-1769-1640}
Peter Doel,$^{12}$
Ni Putu Audita Placida Emas,$^{3}$
Simone Ferraro,$^{9}$\orcidlink{0000-0003-4992-7854}
Jaime E. Forero-Romero,$^{28}$\orcidlink{0000-0002-2890-3725}
Cristhian Garcia-Quintero,$^{14}$\orcidlink{0000-0003-1481-4294}
Enrique Gaztañaga,$^{15}$
Satya Gontcho A Gontcho,$^{9}$\orcidlink{0000-0003-3142-233X}
Boryana Hadzhiyska,$^{9}$\orcidlink{0000-0002-2312-3121}
Sven Heydenreich,$^8$
Klaus Honscheid,$^7$
Mustapha Ishak,$^{14}$\orcidlink{0000-0002-6024-466X}
Shahab Joudaki,$^{16}$\orcidlink{0000-0001-8820-673X}
Eric Jullo,$^{17}$\orcidlink{0000-0002-9253-053X}
Theodore Kisner,$^9$\orcidlink{0000-0003-3510-7134}
Anthony Kremin,$^9$\orcidlink{0000-0001-6356-7424}
Andrew Lambert,$^9$
Martin Landriau,$^{9}$\orcidlink{0000-0003-1838-8528}
Marc Manera,$^{18}$\orcidlink{0000-0003-4962-8934}
Aaron Meisner,$^{19}$\orcidlink{0000-0002-1125-7384}
Ramon Miquel,$^{18}$
Jundan Nie,$^{20}$\orcidlink{0000-0001-6590-8122}
Nathalie Palanque-Delabrouille,$^{9}$\orcidlink{0000-0003-3188-784X}
Claire Poppett,$^{9}$
Anna Porredon,$^{21}$\orcidlink{0000-0002-2762-2024}
Mehdi Rezaie,$^{22}$\orcidlink{0000-0001-5589-7116}
Ashley J. Ross,$^7$
Graziano Rossi,$^{23}$
Rossana Ruggeri,$^{3}$\orcidlink{0000-0002-0394-0896}
Eusebio	Sanchez,$^{24}$\orcidlink{0000-0002-9646-8198}
Christoph Saulder,$^{25}$\orcidlink{0000-0002-0408-5633}
Hee-Jong Seo,$^{26}$\orcidlink{0000-0002-6588-3508}
Joseph Harry Silber,$^{9}$\orcidlink{0000-0002-3461-0320}
Gregory Tarlé,$^6$\orcidlink{0000-0003-1704-0781}
Mariana Vargas-Maga\~na,$^{27}$\orcidlink{0000-0003-3841-1836}
Benjamin Alan Weaver,$^{19}$ 
Enia Xhakaj,$^8$
Zhimin Zhou$^{20}$\orcidlink{0000-0002-4135-0977},
and Hu Zou$^{20}$\orcidlink{0000-0002-6684-3997}
}
\vspace{0.3cm}
\\
\parbox{\textwidth}{
The authors' affiliations are shown in Appendix \ref{sec:affiliations}}.
\vspace{-0.5cm}}

\date{Accepted XXX. Received YYY; in original form ZZZ}

\pubyear{2015}

\begin{document}
\label{firstpage}
\pagerange{\pageref{firstpage}--\pageref{lastpage}}
\maketitle
\begin{abstract}
Galaxy--galaxy lensing (GGL) and clustering measurements from the Dark Energy Spectroscopic Instrument Year 1 (DESI Y1) dataset promise to yield unprecedented combined-probe tests of cosmology and the galaxy--halo connection. In such analyses, it is essential to identify and characterise all relevant statistical and systematic errors. We forecast the covariances of DESI Y1 GGL+clustering measurements and the systematic bias due to redshift evolution in the lens samples. Focusing on the projected clustering and GGL correlations, we compute a Gaussian analytical covariance, using a suite of N-body and log-normal simulations to characterise the effect of the survey footprint. Using the DESI One Percent Survey data, we measure the evolution of galaxy bias parameters for the DESI Luminous Red Galaxy (LRG) and Bright Galaxy Survey (BGS) samples. We find mild evolution in the LRGs in $0.4 < z < 0.8$, subdominant to the expected statistical errors. For BGS, we find less evolution for brighter absolute magnitude cuts, at the cost of reduced sample size. We find that for a redshift bin width $\Delta z = 0.1$, evolution effects on DESI Y1 GGL is negligible across all scales, all fiducial selection cuts, all fiducial redshift bins. Galaxy clustering is more sensitive to evolution due to the bias squared scaling. Nevertheless the redshift evolution effect is insignificant for clustering above the 1-halo scale of $0.1h^{-1}$Mpc. For studies that wish to reliably access smaller scales, additional treatment of redshift evolution is likely needed. This study serves as a reference for GGL and clustering studies using the DESI Y1 sample.
\end{abstract}



\section{Introduction}

The large-scale structure (LSS) in our Universe carries a wealth of information, and has become an essential probe in constraining fundamental physics and testing the current cosmological paradigm and galaxy formation.  Accessing the information contained in the LSS requires summary statistics that capture the salient features in the observed density field. The galaxy auto-correlation function has long been the workhorse summary statistic, and has produced some of the earliest LSS constraints on the underlying cosmological model \citep[e.g.][]{2001Percival, 2002Hamilton, 2005Cole, 2006Tegmark, 2007Padmanabhan}. In the last two decades, weak gravitational lensing has emerged as another powerful probe of cosmology and gravity that complements the information content of the galaxy auto-correlation function \citep[e.g.][]{2023Dalal, 2023Lange, 2022des, 2022Leauthaud, 2021Heymans, 2018Mandelbaum, 2016Blake, 2010Schrabback}. 

Weak gravitational lensing refers to the correlated gravitational distortion induced in background galaxy shapes by foreground LSS, as their light travels towards us. This effect is a powerful cosmological probe because it is sensitive to the geometry along the light path and the potential wells of the intervening structure. In principle, we can extract the cosmological information through auto-correlation of the observed shapes of galaxies, commonly referred to as cosmic shear (see \citealt{2015Kilbinger} for a review), or by cross-correlation of the shapes of the galaxies with positions of foreground galaxies, often referred to as galaxy-galaxy lensing \citep[GGL; e.g.][]{2013Mandelbaum, 1996Brainerd}. In this paper, we are primarily concerned with the combination of galaxy-galaxy lensing with galaxy auto-correlation functions, which improves the constraining power on cosmology by breaking degeneracies with galaxy bias \citep[e.g.][]{2023Dvornik, 2022Pandey, 2019Yuan, 2017Leauthaud, 2015Coupon}. 

The Dark Energy Spectroscopic Instrument (DESI) is a spectroscopic galaxy survey with the primary goal of determining the nature of dark energy through the most precise measurement of the expansion history of the universe ever obtained \citep{2022Abareshi, 2016DESI, 2013Levi}. The baseline survey began in 2021 and will obtain spectroscopic measurements of 40 million galaxies and quasars in a 14,000 deg$^2$ footprint in five years. This represents an order-of-magnitude improvement both in the volume surveyed and the number of galaxies measured over previous surveys such as BOSS and eBOSS \citep{2020Alam, 2020Ahumada}. The DESI large-scale structure samples are divided into 4 target classes: the bright galaxy survey (BGS) sample, the luminous red galaxies (LRG), the emission line galaxies (ELG), and the quasi-stellar objects (QSO), with increasingly high redshift kernels. The LRG and BGS samples at lower redshift provide a clean and excellently calibrated foreground lens sample for galaxy-galaxy lensing studies. With these two samples, DESI will provide unprecedentedly high signal-to-noise measurement of galaxy-galaxy lensing over large overlap footprints with photometric surveys such as Hyper-Suprime Cam (HSC) survey \citep[][]{2018Aihara, 2018Miyazaki, 2018aMandelbaum, 2018bMandelbaum}, the Dark Energy Survey \citep[DES,][]{2016DES, 2016Park, 2018Drlica}, the Kilo-Degree Survey \citep[KiDS-1000,][]{2019Kuijken, 2020Wright, 2021Hildebrandt, 2021Giblin, 2021Heymans}, and in the future, Euclid space telescope \citep{2020Euclid, 2011Laureijs} and Vera Rubin Observatories' Legacy Survey of Space and Time \citep[LSST,][]{2008Ivezic}.

In order to robustly analyze the lensing and clustering measurements to the necessary precision in DESI, we need to understand and characterize all the relevant systematics, and accurately determine the covariance between the measured correlations.  A leading systematic error arises when the observed lens galaxy samples evolve over the width of the redshift bin, but the model ignores such evolution and makes predictions at a fixed redshift. This simplification rides on the assumption that the targeted sample is sufficient uniform within a redshift bin, but a significantly evolving galaxy bias can break this assumption. Thus, when selecting a lens sample for cosmological analysis, it is essential to test this assumption and characterize any potential biases. 
Several studies have leveraged existing data to demonstrate significant redshift evolution in the galaxy bias of LRGs \citep{2017Zhai, 2019bLange, 2023Yuan, 2023Pandey}. Specifically, \cite{2023Pandey} found that ignoring the evolution can lead to approximately 1$\sigma$ biases in the cosmological constraints in future Rubin LSST analyses. 


In this paper, we model the redshift evolution of the DESI BGS and LRG samples using the DESI One-Percent survey clustering measurements \citep{sv, edr}. From our best-fit models, we predict the systematic biases due to redshift binning and compare them to the statistical uncertainties of DESI Y1. We note several key trade-offs: (1) increasing the number of redshift bins reduces the evolution effect within each bin but results in noisier covariances; (2) a higher absolute magnitude cut results in a more uniform sample but at the cost of reduced sample size; (3) a higher maximum redshift in the lens sample increases the sample size but at the cost of increased intrinsic alignment (IA) contamination.  Finally, we also address the needs of analyses that rely on pre-determined simulation snapshots for modeling. This paper accompanies \cite{DESI_Lange}, which assesses additional theoretical systematic effects on the DESI GGL measurements.

We also present our determination of the analytical covariance for DESI Y1 analyses of projected galaxy-galaxy lensing and clustering statistics.  This error forecast is required for both assessing the importance of systematic errors, such as due to lens evolution, as well as performing the future cosmological analysis.  Our fiducial covariance is calculated using an analytical approach, following previous treatments by (e.g.) \cite{2017MNRAS.471.3827S, 2018MNRAS.478.4277S, 2018MNRAS.479.1240D, 2020A&A...642A.158B}.  In this paper we use N-body and log-normal simulations, sampling the same footprint as the data, to test and improve the efficacy of this approach.

This paper is organized as follows. In section~\ref{sec:data}, we describe the DESI LRG and BGS samples and the photometric source samples used for this analysis.  We present the analytic covariance matrix calculation in section~\ref{sec:cov}, and in section~\ref{sec:method} we describe the pipeline we use to model the redshift evolution of the lens samples. In section~\ref{sec:results}, we predict the lens bin biases of the DESI Y1 LRG and BGS samples compared to the statistical noise for a fiducial set of redshift bins. We summarise our findings in section~\ref{sec:conclude}.

\section{Data}
\label{sec:data}

The fiducial DESI Y1 galaxy-galaxy lensing analysis will utilize the LRG and BGS samples as lenses, and DES, KiDS and HSC samples as sources.
We briefly describe these samples in this section. We also introduce the relevant summary statistics for Y1 analysis: galaxy clustering and galaxy-galaxy lensing.

To model the redshift evolution of the lens samples, we make use of the DESI One Percent Survey data. DESI observed its One Percent Survey as the third and final phase of its science validation (SV) program in April and May of 2021. Observation fields were chosen to be in 20 non-overlapping `rosettes' selected to cover major datasets from other surveys, including the Cosmic Evolution Survey (COSMOS), HSC, DES deep field, Galaxy And Mass Assembly (GAMA), Great Observatories Origins Deep Survey (GOODS), and anticipated deep fields from future LSST and Euclid observations. Each rosette is observed at least 12 times, resulting in over $99\%$ completeness for the LRG and BGS samples.  The One Percent Survey essentially produces a smaller but more complete preview version of the upcoming DESI main sample, and is ideal for calibrating galaxy-halo connection models and building high fidelity mocks.  Full details of this sample are presented in \cite{sv} and \cite{edr}.

\begin{figure*}
    \hspace{-0.7cm}
    \includegraphics[width=0.9\textwidth]{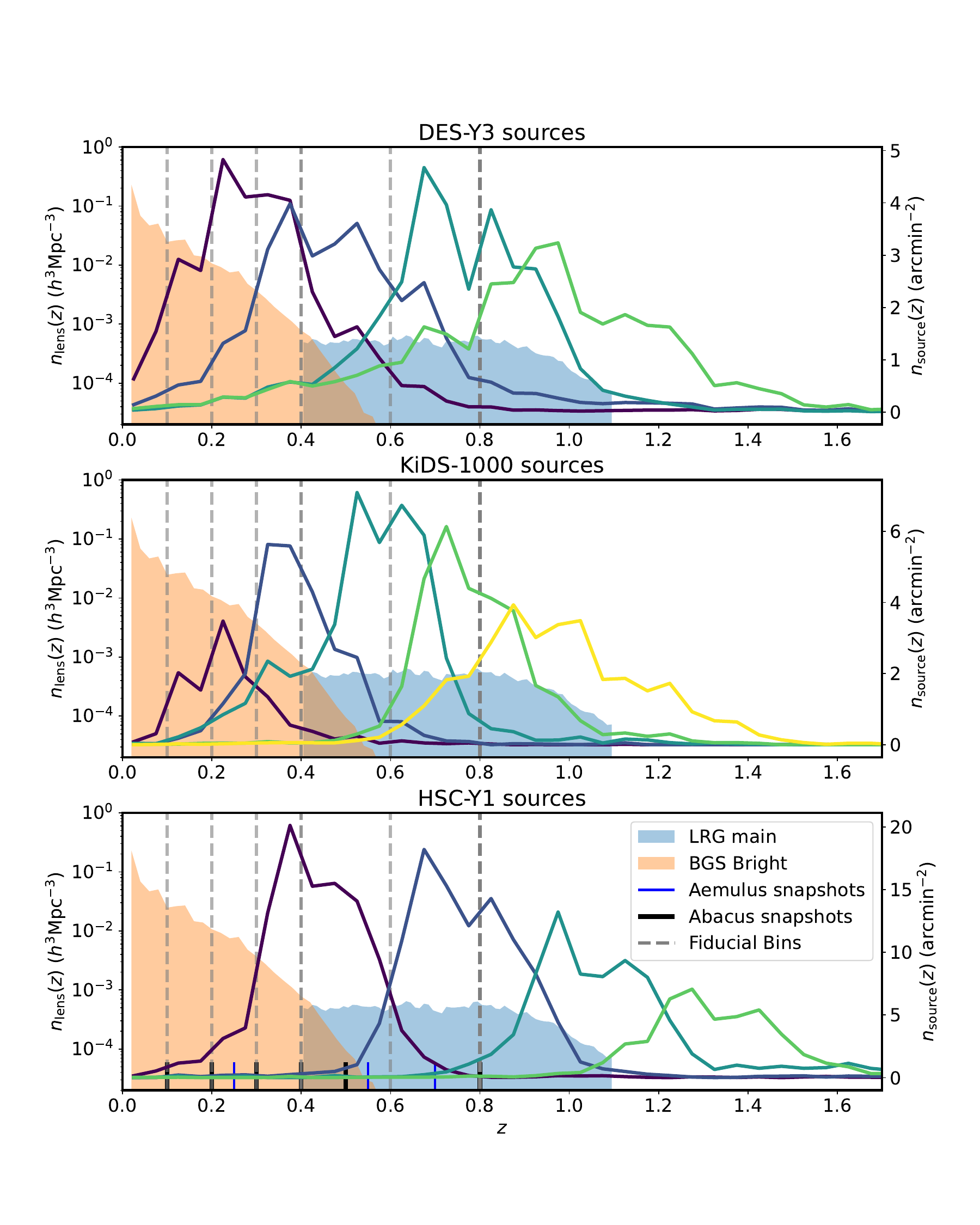}
    \vspace{-1.5cm}
    \caption{The DESI lens density-redshift distribution as seen in the One-percent sample. We also over-plot the source galaxy distribution in each tomographic bin. The three panels correspond to three source galaxy datasets, where the grey vertical dashed lines represent the default lens bins. The short solid vertical lines in the bottom panel mark the \textsc{AbacusSummit} and \textsc{Aemulus} simulation snapshots for the reference. }
    \label{fig:lens_nz}
\end{figure*}

\subsection{DESI LRG}

LRGs are essential for large-scale structure studies. LRGs are specifically selected for observations due to two main advantages: 1) they are bright galaxies with the prominent 4000 \AA\ break in their spectra, thus allowing for relatively easy target selection and redshift measurements; and 2) they are highly biased tracers of the large-scale structure, thus yielding a higher S/N per-object for clustering measurements compared to typical galaxies. The LRG SV target selection is defined in \citet{2020Zhou}. The DESI LRG sample has a target density of 605 deg$^{-2}$ in the redshift range $0.4 < z < 0.8$, significantly higher than previous LRG surveys \citep[BOSS and eBOSS][]{2013Dawson, 2016Dawson}, while the sample also extends to $z \sim 1$. Within the DESI One Percent Survey, the LRG main sample consists of 89,059 galaxies, 43,269 in the northern footprint and 45,790 in the southern footprint. The shaded blue histogram in Figure~\ref{fig:lens_nz} shows the mean spatial density of the LRG sample as a function of redshift. We only consider the LRG sample in the redshift interval $0.4 < z < 0.8$ for fiducial cosmology analyses. 

\subsection{DESI BGS}

\begin{figure}
    \includegraphics[width=0.45\textwidth]{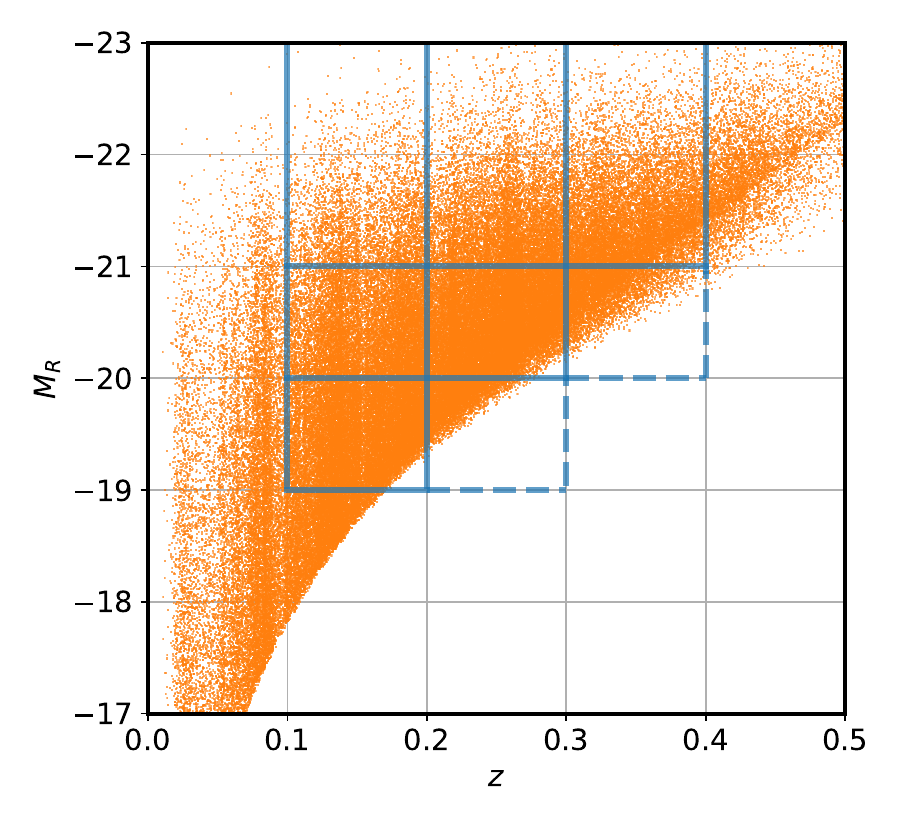}
    \vspace{-0.4cm}
    \caption{The distribution of the DESI One-percent survey BGS Bright sample in $r$-band absolute magnitude and redshift. The thick lines indicate the redshift bins and magnitude cuts considered in this paper. Two bins are outlined with dashed borders to indicate that these bins are highly incomplete. }
    \label{fig:bgs_mrz}
\end{figure}

During DESI bright time, a BGS sample is observed, which has a `Bright' and `Faint' component, as well as Milky Way stars. In this work, we only consider the higher priority BGS Bright sample for simplicity and do not consider the faint sample as it suffers from complicated incompleteness and systematic effects. The BGS SV target selection is defined in \cite{2023Hahn, 2021MNRAS.502.4328R}.  Within the One Percent Survey, the BGS Bright sample consists of 115,602 galaxies in redshift range $0.1 < z < 0.4$. The mean number density as a function of redshift is shown through the orange shaded histograms in Figure~\ref{fig:lens_nz}, whereas Figure~\ref{fig:bgs_mrz} shows the full distribution of BGS Bright galaxies in the absolute magnitude-redshift plane. The thick blue lines show the fiducial selection cuts we adopt for this analysis. We have three $R$-band magnitude cuts at -19, -20, -21, and three redshift bins 0.1-0.2, 0.2-0.3, 0.3-0.4. The dashed lines show selections that are highly incomplete: $M_R < -19$ in $0.2 < z < 0.3$, and $M_R < -20$ in $0.3 < z < 0.4$. These cuts are sub-optimal as they do not provide meaningful statistical gains over brighter more complete selections in the same redshift bins. Thus, these two selections are likely not relevant for our final analysis setup. We include these bins in our results for completeness. 

\begin{figure}
    \includegraphics[width=0.45\textwidth]{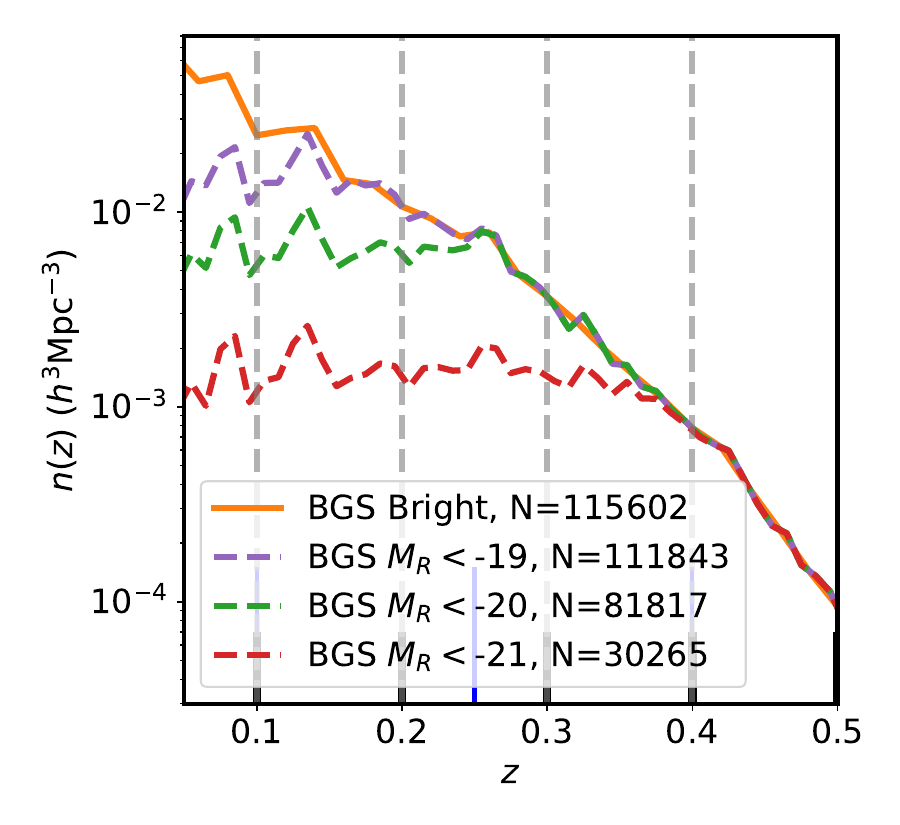}
    \vspace{-0.4cm}
    \caption{The DESI BGS mean density as a function of redshift for different magnitude cuts. The legend also states the BGS sample size for different magnitude cuts, within the fiducial redshift range of $0.1 < z < 0.4$.}
    \label{fig:lens_nz_bgs}
\end{figure}

Given that the BGS is an apparent-magnitude selected sample, a key consideration when attempting to define a homogeneous subset of lens galaxies is the absolute magnitude cut. For the DESI fiducial large-scale BGS clustering analyses, a conservative magnitude cut of $M_R < -21.5$ is sufficient, as it renders the Poisson noise negligible on the large scales relevant to Baryon Acoustic Oscillation (BAO) analysis. However, given galaxy-galaxy lensing analyses are noise-limited rather than sample variance-limited, we intend to retain a much larger fraction of the BGS sample to maximize statistical precision. Figure~\ref{fig:lens_nz_bgs} shows how different absolute magnitude cuts change the mean number density of the BGS sample as a function of redshift. Applying a $M_R < -19$ cut removes approximately $3.3\%$ of the sample; a $M_R < -20$ cut removes approximately 31$\%$ of the sample; $M_R < -21$ cut removes approximately 76$\%$ of the sample. The absolute magnitudes are defined in Eq.~6 in \cite{edr}.  Figure~\ref{fig:lens_nz_bgs} shows the mean number density as a function of redshift for these different absolute magnitude cuts. A brighter magnitude cuts results in a more uniform $n(z)$, but at the cost of removing a large fraction of the sample. We will analyse the effects of these choices on the resulting statistical and systematic errors in this study.

\subsection{Photometric sources}

In this study we consider a cross-correlation analysis of these DESI lens samples with three weak lensing datasets with significant overlap with DESI: the DES Year 3 dataset \citep[DES-Y3][]{2021Gatti}, the current KiDS dataset \citep[KiDS-1000][]{2021Giblin} and the HSC Year 1 dataset \citep[HSC-Y1][]{2018aMandelbaum}.

The Dark Energy Survey has used the Dark Energy Camera at the Blanco 4-m telescope to image 5000 deg$^2$ in $grizY$ filters.  The Y3 sample comprises data taken during the first three years of DES operations, covering the full footprint with approximately half the final exposure time, resulting in a catalogue with depth $i \approx 23$.  The DES-Y3 shear catalogue contains approximately 100 million galaxies covering an area of 4143 deg$^2$, with effective source number density $n_{\rm eff} = 5.6$ arcmin$^{-2}$ and shape noise $\sigma_e = 0.27$.  The sample is divided for tomographic analysis into 4 photometric redshift bins with median redshifts $z = [0.34, 0.52, 0.74, 0.96]$ \citep{2022PhRvD.105b3514A}, whose redshift distributions are calibrated by self-organising maps (SOMs) and further constrained by clustering cross-correlations \citep{2022Gatti,2021Myles, 2019Buchs}.  We adopt these redshift distributions in our forecasts described below.

The Kilo-Degree Survey is a European Southern Observatory public survey using a combination of optical imaging in $ugri$ bands from the 2.6-m VLT Survey Telescope with depth $r \approx 25$, and overlapping near-infrared imaging in $ZYJHK_s$ bands from the VISTA-VIKING survey.  The KiDS-1000 lensing catalogue is based on the fourth KiDS data release, comprising 1006 deg$^2$ of imaging data.  The sample consists of 21 million galaxies with $n_{\rm eff} = 6.2$ arcmin$^{-2}$ and $\sigma_e = 0.27$ \citep{2021Giblin}, divided into 5 photometric bins for tomographic analysis by photometric redshift $z_B$ derived using the BPZ method, with bin limits $z_B = [0.1, 0.3, 0.5, 0.7, 0.9, 1.2]$.  The redshift distributions of these sources samples were determined by the KiDS collaboration using the SOM methodology \citep{2021Hildebrandt}.

Finally, Hyper Suprime-Cam is a wide-field imaging camera on the 8.2-m Subaru telescope, which has been used to conduct a 6-year deep, multi-band imaging survey.  The Wide layer, which is designed for weak lensing cosmology, aims at covering 1400 deg$^2$ in $grizy$ bands with depth $r \approx 26$.  The HSC first-year shear catalogue is based on about 90 nights of HSC Wide data covering 137 deg$^2$ with depth $i \approx 24.5$, and contains about 9 million galaxies with $n_{\rm eff} = 17.6$ arcmin$^{-2}$ \citep{2018aMandelbaum}.  The source sample is split into tomographic bins by photometric redshifts with bin limits $z = [0.3, 0.6, 0.9, 1.2, 1.5]$, with the redshift distributions for each bin estimated using the COSMOS 30-band photo-$z$ catalogue.

\subsection{Summary statistics}

For this analysis we consider two summary statistics: the projected 2-point galaxy correlation function (2PCF), $w_p$, and the surface mass density contrast, $\Delta\Sigma$, measured by galaxy-galaxy lensing.  We start by introducing the 3D 2PCF, which can be compressed into a 2D function due to rotational symmetry. We can express this 2D correlation function in terms of transverse separation $r_p$ and line-of-sight (LoS) separation $r_\pi$, $\xi(r_p, r_\pi)$, which we can compute using the \citet{1993Landy} estimator:
\begin{equation}
    \xi(r_p, r_\pi) = \frac{DD - 2DR + RR}{RR},
    \label{equ:xi_def}
\end{equation}
where $DD$, $DR$, and $RR$ are the normalized numbers of data-data, data-random, and random-random pair counts in each bin of $(r_p, r_\pi)$. The redshift-space $\xi(r_p, r_\pi)$ represents the full information content of the 2PCF.  

To avoid the complexities of modeling the small-scale finger-of-god effect, we can further compress $\xi(r_p, r_\pi)$ to the projected galaxy 2PCF, $w_p$, which is the line-of-sight integral of $\xi(r_p, r_\pi)$,
\begin{equation}
w_p(r_p) = 2\int_0^{r_{\mathrm{\pi, max}}} \xi(r_p, r_\pi) \, dr_\pi .
\label{equ:wp_def}
\end{equation}
Whilst $w_p$ is strictly less informative than $\xi(r_p, r_\pi)$ as it excludes the velocity information encoded in the LoS clustering, it is easier to model as it avoids non-linear velocity effects, and its covariance matrix is more tractable owing to the data compression. It is common for combined-probe analyses to consider $w_p$ in addition to weak lensing measurements for breaking of parameter degeneracies, for example associated with galaxy bias.

The galaxy-galaxy lensing observable we use is the mean surface mass density contrast profile $\Delta \Sigma$, defined as
\begin{equation}
    \Delta\Sigma(r_p) = \overline{\Sigma}(<r_p) - \overline{\Sigma}(r_p),
\label{equ:ds_def}
\end{equation}
where $\overline{\Sigma}(r_\perp)$ is the azimuthally averaged and projected surface mass density at radius $r_p$ and $ \overline{\Sigma}(<r_p)$ is the mean projected surface mass density within projected separation $r_p$ \citep{1991Miralda-Escude, 2001Wilson, 2017Leauthaud}:  
\begin{equation}
\label{eqn:apmd}
\overline{\Sigma}(\leq r_p)=\frac{2}{r_p^2} \int^{r_p}_0 \overline{\Sigma}(R') R' \rm{d}R' \, .
\end{equation}




\section{Analytical Covariances}
\label{sec:cov}

\subsection{Computation of analytical covariance for projected correlations}

The assessment of the significance of systematics due to lens evolution, as well as the statistical analysis of the survey measurements, depends on the covariance of the measured correlation functions, (for example) $\Delta \Sigma_i$ and $\Delta \Sigma_j$ for lens-source combinations $i$ and $j$ between two scales $r_{p,k}$ and $r_{p,l}$,
\begin{equation}
\begin{split}
    & {\rm Cov} \left[ \Delta \Sigma_i(r_{p,k}) , \Delta \Sigma_j(r_{p,l}) \right] = \\
    & \hspace{5mm} \langle \Delta \Sigma_i(r_{p,k}) \, \Delta \Sigma_j(r_{p,l}) \rangle - \langle \Delta \Sigma_i(r_{p,k}) \rangle \, \langle \Delta \Sigma_j(r_{p,l}) \rangle .
\end{split}
\end{equation}
We use analytical methods to determine the covariances between different statistics, redshifts and scales for the DESI BGS and LRG samples and the DES, KiDS and HSC weak lensing datasets. Our covariance computation follows the methods of \cite{2020A&A...642A.158B} (see also \cite{2017MNRAS.471.3827S, 2018MNRAS.478.4277S, 2018MNRAS.479.1240D}) and is fully described in Appendix \ref{sec:covformulae}; we provide a brief summary here.

We compute a Gaussian analytical covariance including the sample variance, noise and mixed contributions.  We assume a fiducial non-linear matter power spectrum, the measured DESI galaxy redshift distributions in each bin, and the weak lensing source survey configurations in tomographic bins defined for DES-Y3 \citep{2022PhRvD.105b3514A}, KiDS-1000 \citep{2021Giblin} and HSC-Y1 \citep{2019PASJ...71...43H} including the shape noise and effective source number density in each bin.  We evaluate the covariance in 15 logarithmic bins of projected separation in the range $0.08 < r_p < 80 \, h^{-1}$ Mpc.  For the $w_p$ covariance, we set $r_{\pi,{\rm max}} = 100 \, h^{-1}$ Mpc.  We assume the linear bias factors determined for early DESI data by \cite{2023Prada}.  The covariance is noise-dominated on small scales, so the assumption of linear bias does not have a significant impact on these scales.  We evaluate the covariances for the forecast overlap areas of DESI-Y1 and these weak lensing surveys: (716, 456, 142) deg$^2$ for DES-Y3, KiDS-1000 and HSC-Y1 overlapping with BGS, and (845, 455, 153) deg$^2$ with LRGs.

\subsection{Covariance comparisons with N-body simulations and jack-knife errors}

We tested the analytical covariance determination for the surface mass density contrast $\Delta\Sigma(r_p)$ using the Buzzard DESI-Lensing simulations \citep{DESI_Buzzard}.  These simulations build realistic DESI target and weak lensing source populations within the Buzzard simulation suite \citep{2019arXiv190102401D}.  The DESI targets are included via a halo occupation distribution prescription, and the sources are populated by a statistical method which includes photometric redshift errors, magnitude distributions, source weights and multiplicative shear calibration bias.  We refer the reader to \cite{DESI_Buzzard} for a full description of these simulations.

\begin{figure*}
\includegraphics[width=1.8\columnwidth]{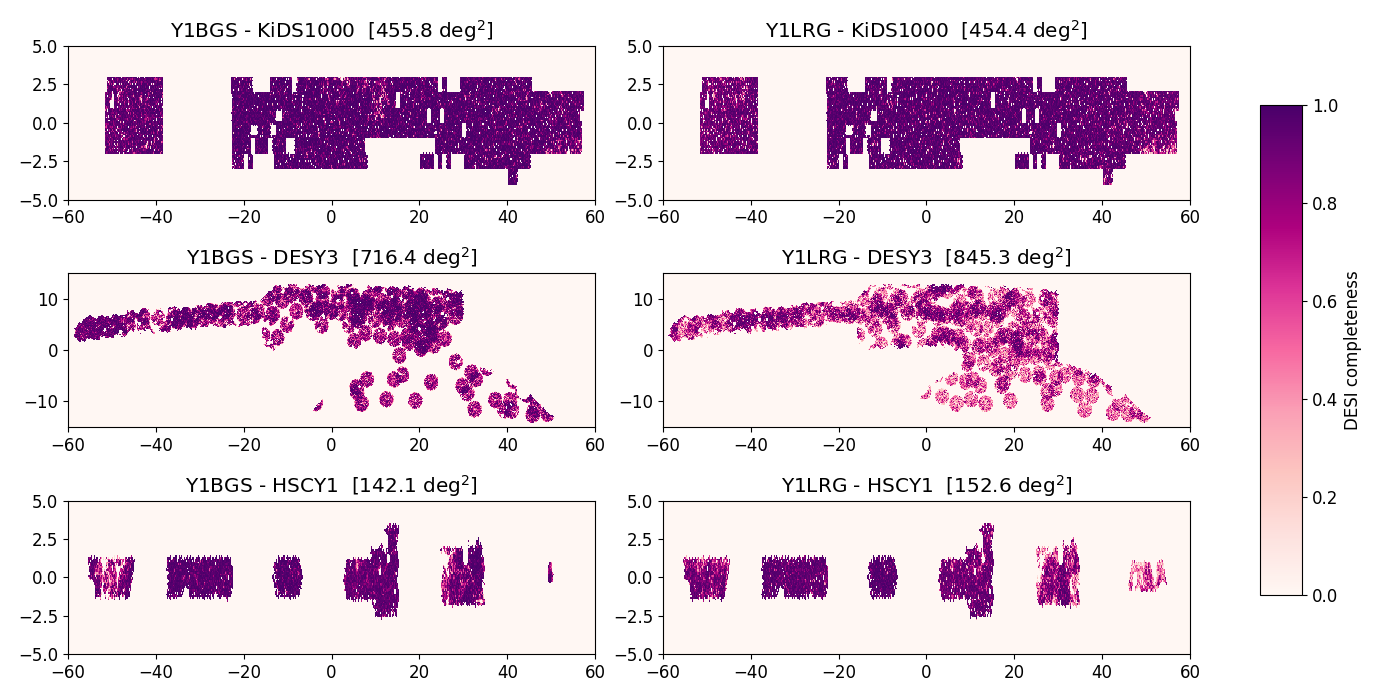}
\caption{The window functions applied when creating the ensemble of Buzzard mocks including survey overlap masks.  Window functions are illustrated for the DESI-Y1 BGS (left-hand column) and LRG observations (right-hand column), using the DESI completeness maps in a {\tt Healpix} $n_{\rm side} = 1024$ pixelisation, where the colour bar indicates the DESI completeness level.  The different rows display the intersection of these DESI completeness maps with the KiDS-1000, DES-Y3 and HSC-Y1 survey footprints.  The axes indicate separation in degrees relative to the field centre, where for the case of HSC-Y1, we have artificially arranged the six widely-separated survey regions in a more compact format.  The areas shown in the panel titles correspond to the total area of intersection between the DESI and lensing survey footprints.}
\label{fig:desiy1_surveys_completeness}
\end{figure*}

We created mock source and lens samples matched to the angular footprint of the overlapping DESI-Y1 and weak lensing datasets.  We determined these ``survey masks'' from the angular completeness maps of the DESI-Y1 redshift catalogues (where the completeness is defined relative to the final DESI target density), intersected with the footprint of each weak lensing dataset, using a {\tt Healpix} $n_{\rm side} = 1024$ pixelisation.  The window functions in each case are illustrated for the three weak lensing surveys in Fig.~\ref{fig:desiy1_surveys_completeness}, where we note that DESI-Y1 BGS and LRG observations have different completeness maps, such that we treat these two redshift ranges separately.  For the case of HSC, we have artificially arranged the six survey regions more compactly in Fig.~\ref{fig:desiy1_surveys_completeness}.  Tiling the irregular geometries of these intersections within a Buzzard quadrant allows (8, 3, 12) realisations to be extracted for the (KiDS, DES, HSC) footprints, producing a total of (64, 24, 96) realisations across the 8 separate Buzzard simulations.

For the different combinations with the source samples, we measured the surface mass density contrast $\Delta\Sigma(r_p)$ around the lenses, using the same measurement code as described in \cite{2020A&A...642A.158B}.  Specifically, we applied the $\Delta\Sigma(r_p)$ estimator assuming a single spectroscopic redshift distribution for each source sample, and therefore we do not utilise the individual source photo-$z$ estimates.  We estimated $\Delta\Sigma(r_p)$ in 15 logarithmically-spaced projected separation bins in the range $0.08 < r_p < 80 \, h^{-1}$ Mpc, and we corrected our results for the multiplicative shear calibration bias introduced in the mocks.  We estimated both the ``tangential'' and ``cross'' components of $\Delta\Sigma(r_p)$.  For the tangential (or ``E-mode'') component, for each source-lens pair included in the estimate, the source shape is projected perpendicular to the source-lens separation vector, measuring the contribution of weak gravitational lensing.  For the cross (or ``B-mode'') component, the source shapes are first rotated by $45^\circ$, which serves as a useful systematic error test which we will apply to the real data.  We refer the reader to \cite{DESI_Y1_GGL_measurements} for a full discussion of $\Delta\Sigma$ estimates in the context of the DESI Y1 data.

\begin{figure*}
\includegraphics[width=1.8\columnwidth]{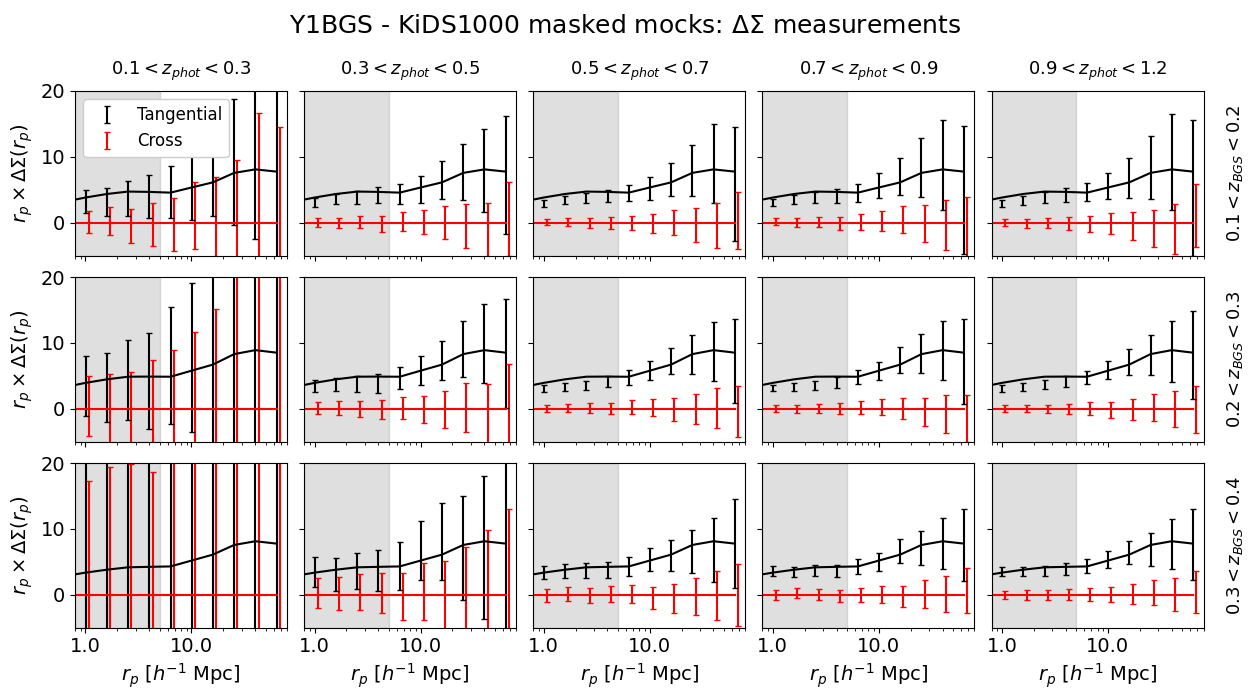}
\caption{The ``tangential'' and ``cross'' components of the surface mass density contrast $\Delta\Sigma(r_p)$ of the five tomographic source samples of the KiDS-1000 mocks, around the three lens samples of the DESI-Y1 BGS mocks, for the simulations including the overlap mask.  The different panels display measurements for different combinations of source and lens samples, as indicated by the captions above and to the right of each panel.  We plot the mean and standard deviation of the measurements across 64 Buzzard masked regions.  The solid lines show the fiducial cosmological models, where a linear bias factor has been jointly fit to each lens redshift slice using the separation range $r_p > 5 \, h^{-1} \, {\rm Mpc}$ (indicated by the un-shaded regions).  The units of $\Delta\Sigma(r_p)$ are $h \, M_\odot \, {\rm pc}^{-2}$, and the $y$-axes are scaled by a factor of $r_p$ (in $h^{-1}$ Mpc) for clarity of presentation.}
\label{fig:ds_Y1BGS_kids}
\end{figure*}

We illustrate our results using the example of the masked simulations of DESI-Y1 BGS and KiDS-1000, noting that the analyses of the other survey configurations reach similar conclusions. Fig.\ref{fig:ds_Y1BGS_kids} displays the tangential and cross $\Delta\Sigma(r_p)$ measurements between the five tomographic source samples of the KiDS-1000 mocks, around the three lens samples of the DESI-Y1 BGS mocks ($0.1 < z < 0.2$, $0.2 < z < 0.3$ and $0.3 < z < 0.4$).  We plot the mean and standard deviations of the measurements across the 64 realisations, compared to the fiducial cosmological models in each case, scaled by a linear galaxy bias factor jointly fit to the data at each lens redshift.  The models are a good description of the measurements, and (in these mock catalogues) we detect no evidence of a ``cross'' component.

\begin{figure*}
\includegraphics[width=1.8\columnwidth]{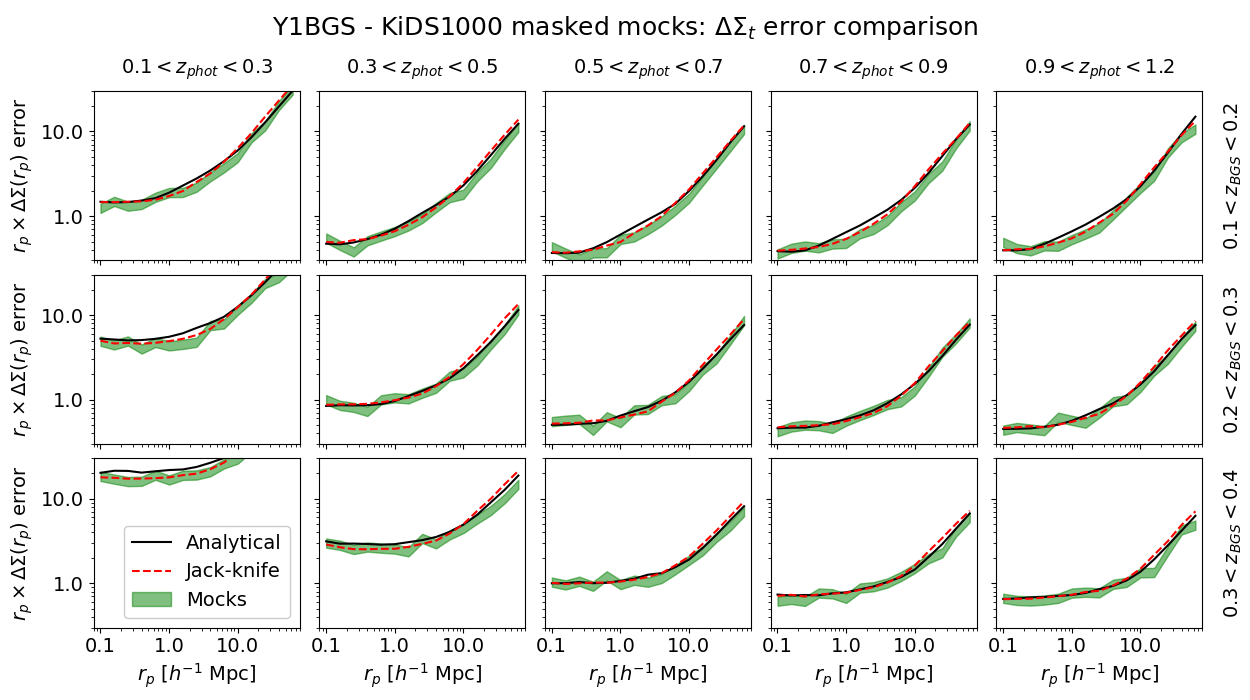}
\caption{A comparison between error estimates of the tangential component of the surface mass density contrast $\Delta\Sigma(r_p)$ of the five tomographic source samples of KiDS-1000, around the three lens samples of DESI-BGS, for the Buzzard mocks including survey overlap masks.  The different panels display error estimates for different combinations of source and lens samples.  We compare the error in $\Delta\Sigma(r_p)$ predicted by the analytical covariance (solid black line), with the standard deviation of the measurements across 64 Buzzard regions (the green band, which indicates the error in the inferred standard deviation arising from the limited number of regions), with the jack-knife error (dashed red line).}
\label{fig:dserr_comparison_Y1BGS_kids}
\end{figure*}

Fig.~\ref{fig:dserr_comparison_Y1BGS_kids} displays a comparison between different estimates of the error in these $\Delta\Sigma$ measurements.  In this figure we compare the estimate from the analytical covariance (solid black line), with the standard deviation across the 64 mock realisations (represented by the green band, whose width indicates the ``error in the error'' due to the limited number of realisations), with the jack-knife error estimated using 100 jack-knife regions containing equal number of lenses (dashed red line).  The jack-knife regions are defined by boundaries of constant right ascension and declination and have average areas between $1.4$ and $8.5$ deg$^2$ depending on the survey.  The analytical covariance is determined from the Gaussian contribution, together with the noise correction using the measured number of source-lens pairs.  Generally speaking, the different error estimates agree within $10-20\%$, depending on scale and redshift.

\begin{figure*}
\includegraphics[width=\columnwidth]{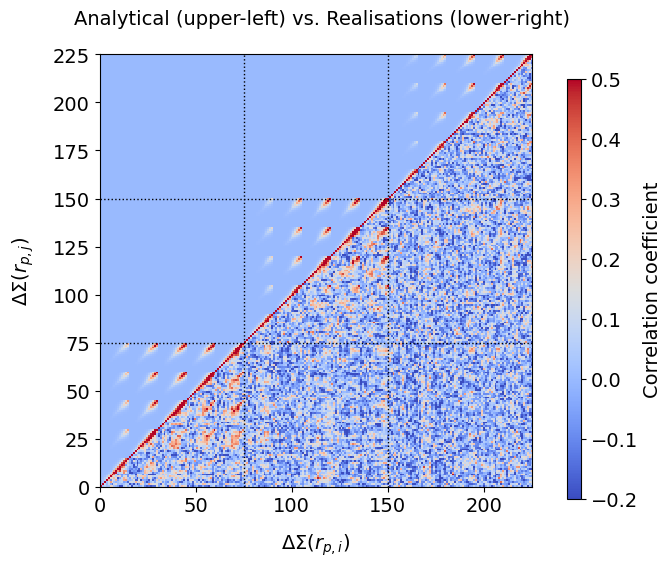}
\includegraphics[width=\columnwidth]{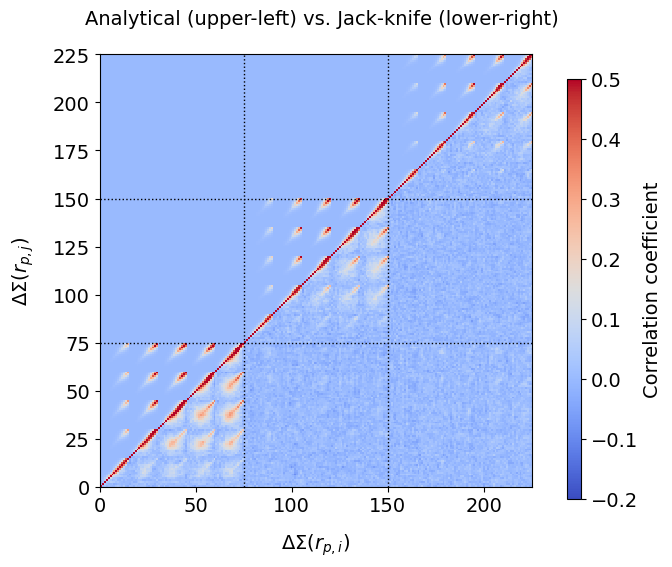}
\caption{The correlation matrix corresponding to the covariance of the tangential $\Delta \Sigma$ measurements of the KiDS-1000 and DESI-Y1 BGS mocks, including the survey masks.  The measurements use $N_{\rm tom} = 5$ source tomographic samples, $N_{\rm lens} = 3$ lens samples and $N_{\rm sep} = 15$ separation bins, which are concatenated into a data vector of length $N_{\rm lens} \cdot N_{\rm tom} \cdot N_{\rm sep} = 225$, where the innermost loop is over separations, and the outermost loop is over lens bins.  The analytical covariance is depicted in the upper-left triangle of both panels.  The lower-right triangles display the numerical covariance derived from 64 realisations (left-hand panel) and the average jack-knife covariance of the 64 realisations (right-hand panel).  The correlation matrix is divided into sections corresponding to the different lens bins, indicated by the horizontal and vertical dotted lines.}
\label{fig:cov_ds_Y1BGS_kids}
\end{figure*}

\begin{figure}
\includegraphics[width=\columnwidth]{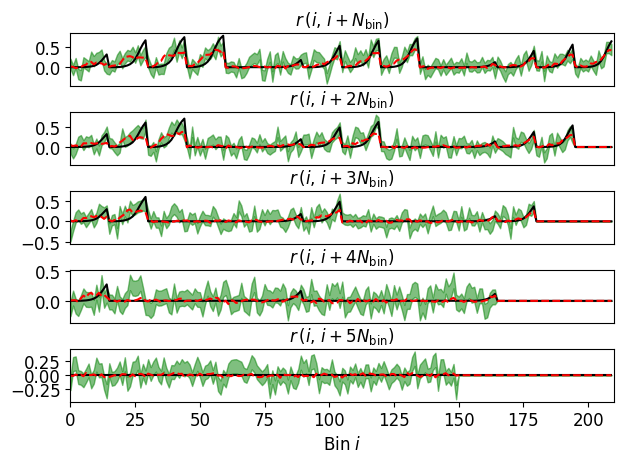}
\caption{The correlation coefficient of the off-diagonal covariance of the tangential $\Delta \Sigma$ measurements of the KiDS-1000 and DESI-Y1 BGS mocks, $r(i,j) = C_{ij}/\sqrt{C_{ii} \, C_{jj}}$.  The different panels display the full set of correlation coefficients for matrix elements sharing the same scale bin, for different lens and source bins offset from the central diagonal.  We compare the correlation coefficient predicted by the analytical covariance (solid black line), the Buzzard realisations (green band) and the jack-knife covariance (dashed red line).}
\label{fig:corr_ds_Y1BGS_kids}
\end{figure}

Fig.~\ref{fig:cov_ds_Y1BGS_kids} displays the full correlation matrix of the $\Delta\Sigma$ data vector, when the different redshift and separation bins are concatenated together, where the analytical covariance is displayed in the upper-left triangle of both panels, and the numerical and jack-knife covariance in the lower-right triangles.  We see a similar qualitative structure in the different estimates of the covariance matrix, which features significant correlation between measurements for different source samples and the same lenses, at adjacent large separations, and negligible correlation between measurements for different lens samples.  Fig.~\ref{fig:corr_ds_Y1BGS_kids} compares estimates of the correlation coefficients for covariance matrix elements sharing the same scale bin, for different lens and source bins offset from the central diagonal.

In Appendix \ref{sec:covfootprint} we use log-normal mocks to investigate the effects of the survey footprint on the galaxy-galaxy lensing covariance.  The window functions of the source and lens samples can alter the sample variance contribution to the Gaussian covariance on large scales \citep{2004A&A...413..465K, 2011ApJ...734...76S, 2019MNRAS.486...52S, 2021MNRAS.508.3125F, 2021A&A...646A.129J}.  We show that these effects are more significant for the cross-component of the galaxy-galaxy lensing, which may be used in a null test for the presence of B-modes.  We find that for the DESI-Y1 footprint geometry shown in Fig.\ref{fig:desiy1_surveys_completeness}, the error in the tangential (cross) component increases with reference to the analytical prediction by $10\%$ ($30\%$) at the scales at which $50\%$ of source-lens pairs are ``lost'' by the boundary.  We describe our corrections to the covariance for these effects in Appendix \ref{sec:covfootprint}.

\section{Modelling lens bin evolution}
\label{sec:method}

In this section we describe how we create our model for the redshift evolution of the DESI lens samples.  Our pipeline first analyzes the DESI One Percent Survey data to constrain the galaxy bias parameters as a function of redshift. Then we produce high-fidelity redshift-dependent mocks using the best-fit galaxy bias model. Finally, we compute the summary statistics and the biases associated with different lens bin choices, whose significance we assess using the analytical covariances.

\subsection{HOD model}
\label{sec:hodmodel}

For galaxy bias modeling, we adopt the Halo Occupation Distribution model (HOD), which probabilistically populates dark matter halos with galaxies according to a set of halo properties. Statistically, the HOD can be summarized as a probability distribution $P(n_g|\boldsymbol{X}_h)$, where $n_g$ is the number of galaxies of the given halo, and $\boldsymbol{X}_h$ is some set of halo properties.

In the vanilla HOD model, halo mass is assumed to be the only relevant halo property $\boldsymbol{X}_h = {M_h}$ \citep{2005Zheng, 2007bZheng}. This vanilla HOD separates the galaxies into central and satellite galaxies, and assumes the central galaxy occupation follows a Bernoulli distribution whereas the satellites follow a Poisson distribution.

For both the LRG and the luminosity-limited BGS samples, the HOD is well approximated by a vanilla model given by \cite{2007bZheng, 2011Zehavi, 2015cGuo}:
\begin{align}
    \bar{n}_{\mathrm{cent}}^{\mathrm{LRG}}(M) & = \frac{f_\mathrm{ic}}{2}\mathrm{erfc} \left[\frac{\log_{10}(M_{\mathrm{cut}}/M)}{\sqrt{2}\sigma}\right], \label{equ:zheng_hod_cent}\\
    \bar{n}_{\mathrm{sat}}^{\mathrm{LRG}}(M) & = \left[\frac{M-\kappa M_{\mathrm{cut}}}{M_1}\right]^{\alpha}\bar{n}_{\mathrm{cent}}^{\mathrm{LRG}}(M),
    \label{equ:zheng_hod_sat}
\end{align}
where the five vanilla parameters characterizing the model are $M_{\mathrm{cut}}, M_1, \sigma, \alpha, \kappa$. We describe these parameters as follows: $M_{\mathrm{cut}}$ sets the minimum halo mass to host a central galaxy; $M_1$ sets the typical halo mass that hosts one satellite galaxy; $\sigma$ controls the steepness of the transition from 0 to 1 in the number of central galaxies; $\alpha$ is the power law index on the number of satellite galaxies; and $\kappa M_\mathrm{cut}$ gives the minimum halo mass to host a satellite galaxy.  We have added a modulation term $\bar{n}_{\mathrm{cent}}^{\mathrm{LRG}}(M)$ to the satellite occupation function to remove satellites from halos without centrals. We have also included an incompleteness parameter $f_\mathrm{ic}$, which is a downsampling factor controlling the overall number density of the mock galaxies. This parameter is relevant when trying to match the observed mean density of the galaxies in addition to clustering measurements. By definition, $0 < f_\mathrm{ic}\leq 1$.

For this analysis, we use the \ahod\ code to find best-fit HODs and sample HOD posteriors. \ahod\ is a highly efficient HOD implementation that enables a large set of HOD extensions \citep[][]{2021bYuan}. The code is publicly available as a part of the \textsc{abacusutils} package at \url{https://github.com/abacusorg/abacusutils}. Example usage can be found at \url{https://abacusutils.readthedocs.io/en/latest/hod.html}. 

The HOD code is implemented on top of the \textsc{AbacusSummit} simulation suite, which is a set of large, high-accuracy cosmological N-body simulations using the \textsc{Abacus} N-body code \citep{2021Maksimova, 2019Garrison, 2021bGarrison}. For this analysis, we use the base simulation box at Planck 2018 cosmology \verb+AbacusSummit_base_c000_ph000+ \citep{2020Planck}. The box contains $6912^3$ particles within a $(2 \, h^{-1}$Gpc$)^3$ volume, which yields a particle mass of $2.1 \times 10^9 \, h^{-1} M_\odot$. \footnote{For more details, see \url{https://abacussummit.readthedocs.io/en/latest/abacussummit.html}}

The dark matter halos are identified with the {\sc CompaSO} halo finder, which is a highly efficient on-the-fly group finder specifically designed
for the \textsc{AbacusSummit} simulations \citep{2021Hadzhiyska}. 
In addition to determining the number of galaxies per halo, the standard HOD model also dictates the position and velocity of the galaxies. In the vanilla model, the position and velocity of the central galaxy are set to be the same as those of the halo center, specifically the L2 subhalo center-of-mass for the {\sc CompaSO} halos. For the satellite galaxies, they are randomly assigned to halo particles with uniform weights, each satellite inheriting the position and velocity of its host particle. 

To extract the redshift evolution, we conduct standard HOD fits in multiple redshift bins centered around \textsc{AbacusSummit} output snapshots. The best-fit parameters are then interpolated as a function of redshift to derive an approximate redshift evolution model, as we describe in the following subsection. 

\subsection{Dark Emulator}


Having obtained the best-fitting HOD parameters in each redshift bin, we interpolate the best-fit parameters as a function of redshift with simple polynomials to build a simple empirical redshift-dependent galaxy bias model.  Then, we apply these models to the Dark Emulator dark matter model to generate realistic redshift-dependent mocks. Dark Emulator essentially provides an emulated model of the halo power spectrum as a function of cosmology, redshift, halo mass, and scale \citep{darkemulator, 2021Kobayashi}. It is trained on a large suite of dark matter only simulations evaluated at 100 different cosmologies. 

Dark Emulator then computes the galaxy correlation and galaxy-galaxy lensing statistics by convolving the vanilla HOD with the emulated halo power spectrum and Fourier transforming back to configuration space. Using Dark Emulator, we evaluate the $w_p$ and $\Delta \Sigma$ statistics along a dense grid of lens redshifts ($\Delta z = 0.01$), with input redshift dependent HOD parameters evaluated from the interpolated HOD$(z)$.

To measure evolution biases from assuming a fixed HOD, we first compute the `real' measurement $ \Delta \Sigma_{\rm evol}$ (or $w_{p, \rm evol}$), which we model by integrating the $\Delta \Sigma$ along the dense redshift grid over the full redshift bin, weighted by the $n(z)$ of the observed sample. Then, we compute the naive model measurement $\Delta \Sigma_{{\rm fixed}}$ (or $w_{p, \mathrm{fixed}}$), which assumes a fixed HOD at a fixed redshift. For analytic theories, this fixed redshift can be approximated as the mean redshift of the sample in the corresponding redshift bin. Thus, $\Delta \Sigma_{{\rm fixed}}$ is computed by simply evaluating our redshift dependent model at the mean redshift. More precisely, 
\begin{align}
    \Delta \Sigma_{{\rm fixed}} & = \Delta \Sigma_{DE} \left( z_\mathrm{mean}, \, {\rm HOD}(z_\mathrm{mean}) \, \right), \\
    \Delta \Sigma_{\rm evol} & = \int dz \, n(z) \, \Delta \Sigma_{DE} \left( z, \, {\rm HOD}(z) \, \right) ,
\end{align}
where $\Delta \Sigma_{DE}$ is the $\Delta \Sigma$ model from Dark Emulator, $n(z)$ is the lens redshift distribution, and $z_\mathrm{mean}$ is the mean redshift in the corresponding redshift bin given $n(z)$. \footnote{We note that assuming mean sample redshift is fairly naive, and additional differences between model and data could be accounted for by shifting the mean redshift. Nevertheless, this assumption is useful in presenting a worst case scenario.} We then use comparisons between these $\Delta \Sigma$ evaluations to determine the impact of these assumptions from different lens bin configurations, characterizing the lens bin bias as, 
\begin{equation}
\delta_\mathrm{z-evol} = \frac{|
\Delta\Sigma_\mathrm{fixed} - \Delta\Sigma_\mathrm{evol}|}{\Delta\Sigma_\mathrm{evol}}.
\label{equ:zbias}
\end{equation}
We can repeat this exercise for $w_p$ to calculate the evolution effect on the projected clustering. We note that assuming $z_\mathrm{mean}$ for the fixed redshift is a conservative scenario, and real analyses may choose to further tune the assumed redshift. 

\section{Systematics due to lens evolution}
\label{sec:results}

In this section, we compare biases due to redshift evolution to the expected statistical error of the DESI Y1 LRG and BGS samples  in a set of fiducial redshift bins.



\subsection{LRG}
To model the redshift evolution of the LRG sample, we conduct HOD fits in the two fiducial redshift bins $0.4 < z < 0.6$ and $0.6 < z < 0.8$. We refer the readers to section~\ref{sec:hodmodel} for the model description and \cite{Yuan2023} for the detailed HOD analysis. We use the posterior means quoted in Table~3 of \cite{Yuan2023} for this analysis. We reproduce the relevant numbers in Table~\ref{tab:lrghods}. 

We build a simple model of LRG redshift evolution HOD$(z)$ by linearly interpolating the best-fit values of $\log M_\mathrm{cut}$ and $\log M_1$ as a function of redshift. The parameters $\alpha$, $\sigma$ and $\kappa$ are poorly constrained and are thus held fixed at their posterior mean values in the lower redshift bin. We then feed the HOD$(z)$ to Dark Emulator to produce a redshift-dependent LRG mock from $z = 0.4$ to $z = 0.8$ and compute the biases in $w_p$ and $\Delta\Sigma$ due to redshift evolution using Eq.~\ref{equ:zbias}.
Because the $n(z)$ of the LRG is fairly flat, the mean redshift of LRGs is fairly close to the middle of the redshift bin. Specifically, we obtain $z_\mathrm{mean} \approx 0.509$ for the $0.4 < z < 0.6$ sample, and $z_\mathrm{mean} \approx 0.705$ for the $0.6 < z < 0.8$ sample. 

We compare the redshift evolution systematic error $\delta_\mathrm{z-evol}$ with the expected statistical error from the same redshift bin, calculated as the cumulative noise-to-signal of the summary statistic, integrating from large to small scales: 
\begin{equation}
    \sigma_\mathrm{stat}(r_p) = \sqrt{\mathbf{x}^T(r > r_p) \, \mathcal{C}^{-1} \, \mathbf{x}(r > r_p)},
\end{equation}
where $\mathbf{x}$ is the target data vector ($w_p$ or $\Delta\Sigma$), and $\mathcal{C}$ is the corresponding analytic covariance matrix forecast in Section~\ref{sec:cov}. One can simply interpret this measure as how accurately we can determine the amplitude of the summary statistics as increasingly smaller scales are incorporated. 

\begin{figure*}
    \hspace{-0.7cm}
    \includegraphics[width=0.8\textwidth]{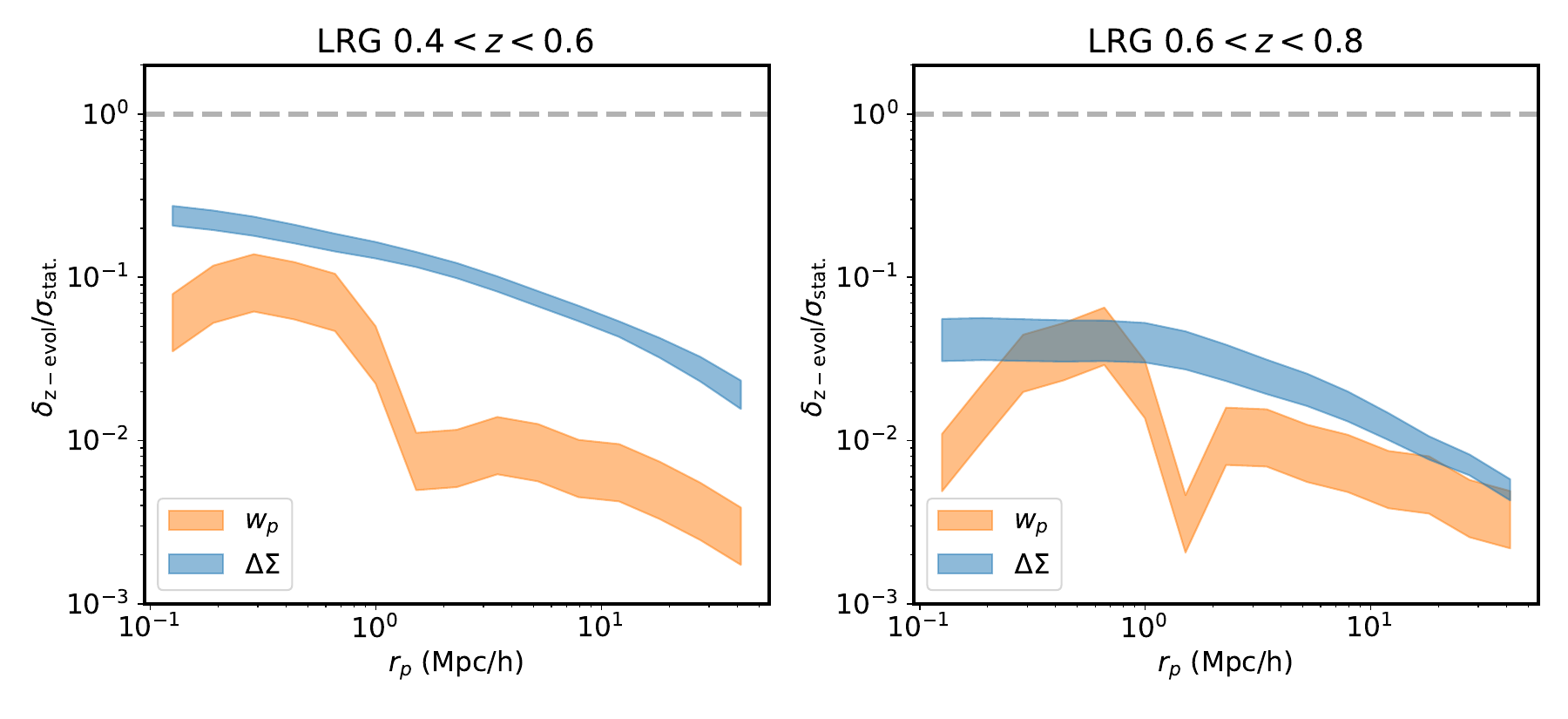}
    \vspace{-0.3cm}
    \caption{The ratio of the error induced by LRG redshift evolution $\delta_\mathrm{z-evol}$ divided by the statistical error $\sigma_\mathrm{stat.}$, as a function of projected scale $r_p$. Note that we show the cumulative measure above a minimum projected scale $r_p$. The two panels correspond to the two fiducial LRG lens bins. The orange curve shows $w_p$, whereas the blue curve shows $\Delta\Sigma$. The width of the two curves indicate the spread between different photometric source samples (DES Y3, KiDS-1000, and HSC). It is clear that for LRGs, the redshift evolution in the fiducial redshift bins is insignificant compared to the statistical error.  }
    \label{fig:lrg_noise_bias}
\end{figure*}

Figure~\ref{fig:lrg_noise_bias} shows the ratio of $\delta_\mathrm{z-evol}$ to the statistical error $\sigma_\mathrm{stat}$ as a function of the minimum projected scale $r_p$ included in the analysis.  We draw a dashed line at $\delta_\mathrm{z-evol}/\sigma_\mathrm{stat} = 1$, above which redshift evolution becomes a significant systematic effect.  The two panels correspond to the two redshift bins, whereas the two colors correspond to the summary statistics $w_p$ and $\Delta\Sigma$. The widths of the band correspond to the spread between different source samples (DES Y3, KiDS-1000, and HSC). It is clear that for LRGs, the systematic due to redshift evolution is insignificant given DESI Y1 statistics. This is true for both fiducial lens bins and for all source samples. This statement also holds down to scales as small as $0.1 \, h^{-1}$Mpc. 





\begin{table} 
    \centering
    {\renewcommand{\arraystretch}{1.5}
    \begin{tabular}{l||c|c}
        \hline
        \hline
        Sample&\multicolumn{2}{c}{$\mathrm{LRG}$}\\[2pt]
        \hline
        $z$ range&
        $0.4<z<0.6$&$0.6<z<0.8$\\
        
        \hline
        $\log M_\mathrm{cut}$&
        12.89&
        12.79
        \\
        
        $\log M_1$&
        14.08&
        13.88
        \\

        $\sigma$&
        0.27&
        0.21
        \\
        
        $\alpha$&
        1.20&
        1.07
        \\
        
        $\kappa$&
        0.65&
        1.4
        \\[2pt]

        \hline
    \end{tabular}
    }
    \caption{LRG HOD posterior means as found in Table~3 of \citet{Yuan2023}. The masses are in units of $h^{-1} M_\odot$.}
    \label{tab:lrghods}
\end{table}

\begin{table*}
    \centering
    {\renewcommand{\arraystretch}{1.5}
    \begin{tabular}{l||c|c|c|c}
        \hline
        \hline
        Sample&\multicolumn{4}{c}{$\mathrm{BGS}\ \  M_R < -19$}\\[2pt]
        \hline
        $z$ range&
        $0.05<z<0.15$&$0.15<z<0.25$&$0.25<z<0.35$&$0.35<z<0.45$\\
        \hline
        $\log M_\mathrm{cut}$
        & 11.59
        & 11.87
        & 12.19
        & -
        \\
        
        $\log M_1$
        & 12.29
        & 12.68
        & 13.14
        & -
        \\

        
        

        \hline

        Sample&\multicolumn{4}{c}{$\mathrm{BGS}\ \  M_R < -20$}\\
        \hline
        $\log M_\mathrm{cut}$
        & 11.89
        & 12.04
        & 12.20
        & 12.39
        \\
        
        $\log M_1$
        & 12.71
        & 12.85
        & 13.15
        & 13.17
        \\

        
        

        \hline

        Sample&\multicolumn{4}{c}{$\mathrm{BGS}\ \  M_R < -21$}\\
        \hline
        $\log M_\mathrm{cut}$
        & 12.59
        & 12.59
        & 12.92
        & 13.12
        \\
        
        $\log M_1$
        & 13.69
        & 13.66
        & 13.50
        & 13.83
        \\

        
        

        \hline
    \end{tabular}%
    }
    \caption{BGS HOD best-fits across 4 redshift bins and 3 absolute magnitude cuts. The best-fit $w_p$ predictions are shown in Figure~\ref{fig:bgs_fit}. Note that we use an non-fiducial redshift range (offset from fiducial bins by 0.05) for the HOD analysis to accommodate the fact that our \textsc{AbacusSummit} simulation snapshots are at $z = 0.1, 0.2, 0.3, 0.4$. We then use these best-fit values to derive the redshift evolution of the sample in the fiducial DESI redshift bins. We have fixed 3 HOD parameters $\sigma = 0.1, \alpha = 0.8, \kappa = 1.0$ for stable fits given the limited S/N in the data vector. We have skipped $M_R < -19$ at $z = 0.4$ since it is the same sample as $M_R < -20$ at $z = 0.4$. The masses are in units of $h^{-1} M_\odot$. These best-fit HODs are also visualized in Figure~\ref{fig:bgs_hod}.
    }
    \label{tab:bgshods}
\end{table*}


\subsection{BGS}

\begin{figure*}
    \hspace{-0.7cm}
    \includegraphics[width=0.85\textwidth]{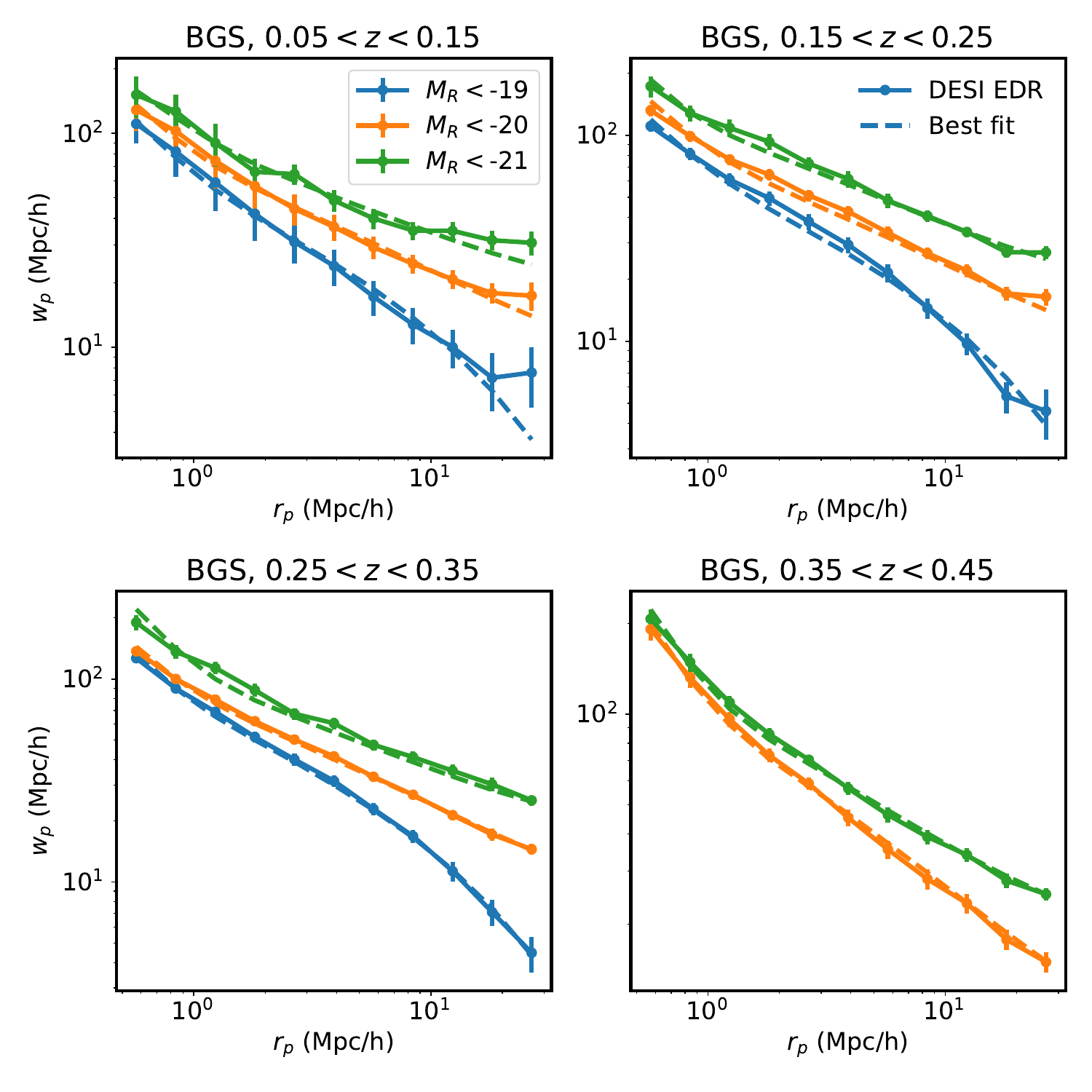}
    \vspace{-0.3cm}
    \caption{The DESI EDR BGS Bright $w_p$ fits across three different absolute magnitude cuts. The solid lines denote the data with jackknife errors. The dashed lines showcase the best-fit prediction. The orange and green curves have been shifted up by $+10$ and $+20$ respectively for visual clarity. Note that we use an non-fiducial redshift range (offset from fiducial bins by 0.05) for the HOD analysis to accommodate the fact that our \textsc{AbacusSummit} simulation snapshots are at $z = 0.1, 0.2, 0.3, 0.4$. We then use these best-fit values to derive the redshift evolution of the sample in the fiducial DESI redshift bins.}
    \label{fig:bgs_fit}
\end{figure*}

For the BGS sample, we perform our analysis not just in terms of redshift binning, but also in terms of cuts in absolute magnitude $M_R$.  We follow the same procedure as for the LRG sample to model the BGS redshift evolution for different magnitude cuts. The BGS sample has not been analyzed with an HOD model in multiple redshift bins in previous papers, so we conduct an additional set of HOD fits of the BGS sample. To construct HOD$(z)$, we conduct vanilla HOD fits of the BGS Bright sample at \textsc{AbacusSummit} snapshots $z = 0.1, 0.2, 0.3, 0.4$, spanning our DESI redshift bins. We repeat the fits for 3 different absolute magnitude cuts, $M_R < -19$, $M_R < -20$, and $M_R < -21$. We only consider the vanilla HOD model for simplicity, and fit the projected 2PCF $w_p$ within $0.1 < r_p < 32 \, h^{-1}\mathrm{Mpc}$. We assume Gaussian likelihoods and use the diagonal errors computed with 60 jackknife regions. The HOD best-fits are summarized in Table~\ref{tab:bgshods}, and the best-fit $w_p$ predictions are displayed in Figure~\ref{fig:bgs_fit}. Note that we have shifted the $w_p$ along the $y$-axis for visual clarity. We have also skipped $M_R < -19$ at $z = 0.4$ since that is the same sample as $M_R < -20$ at $z = 0.4$ (Figure~\ref{fig:bgs_mrz}). Note that we are only varying the two mass parameters in these HOD fits because they are much better constrained than the other parameters given the noisy measurements. 

\begin{figure}
    \hspace{-0.7cm}
    \includegraphics[width=0.5\textwidth]{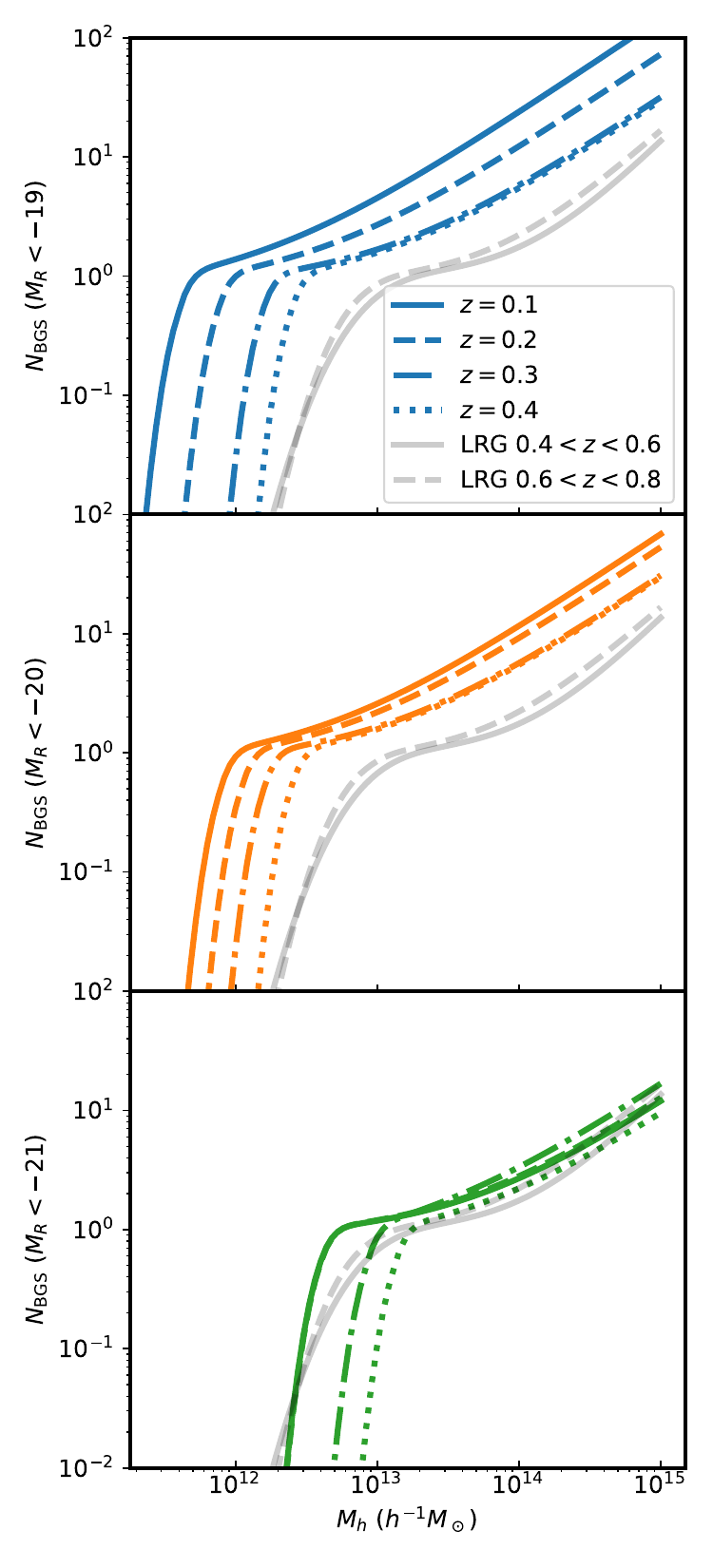}
    \vspace{-0.3cm}
    \caption{The DESI One-percent BGS Bright sample HOD best-fit across three different magnitude-limited subsamples (different panels and colors) and four redshift bins (different line styles). The gray solid and gray dashed lines represent the LRG HOD best-fits from Table~\ref{tab:lrghods}. We see clear redshift evolution in BGS sample. The $M_R < -21$ sample is qualitatively similar to the LRG sample. }
    \label{fig:bgs_hod}
\end{figure}

Figure~\ref{fig:bgs_hod} visualizes the evolution of the best-fit HODs for different magnitude-limited BGS samples. We see clear redshift dependency for all different magnitude cuts. The faintest $M_R < -19$ sample displays the strongest redshift-evolution. The $M_R < -21$ sample is less sensitive to redshift and has a similar HOD to that of the LRG samples shown in gray. The dependence of the mass parameters (and linear bias) on absolute magnitude cuts is consistent with findings of \cite{2023Prada} (see Table~4). The satellite occupation parameters are also similar to that of \cite{2023Prada}. 

We again build an HOD$(z)$ model by interpolating parameters $\log M_\mathrm{cut}$ and $\log M_1$. We compute $\delta_\mathrm{z-evol}$ and $\sigma_\mathrm{stat.}$ for the BGS sample using Dark Emulator, except we repeat these calculations for different magnitude cuts: $M_R < -19$, $M_R < -20$, and $M_R < -21$. 
\begin{figure*}
    \hspace{-0.7cm}
    \includegraphics[width=1.03\textwidth]{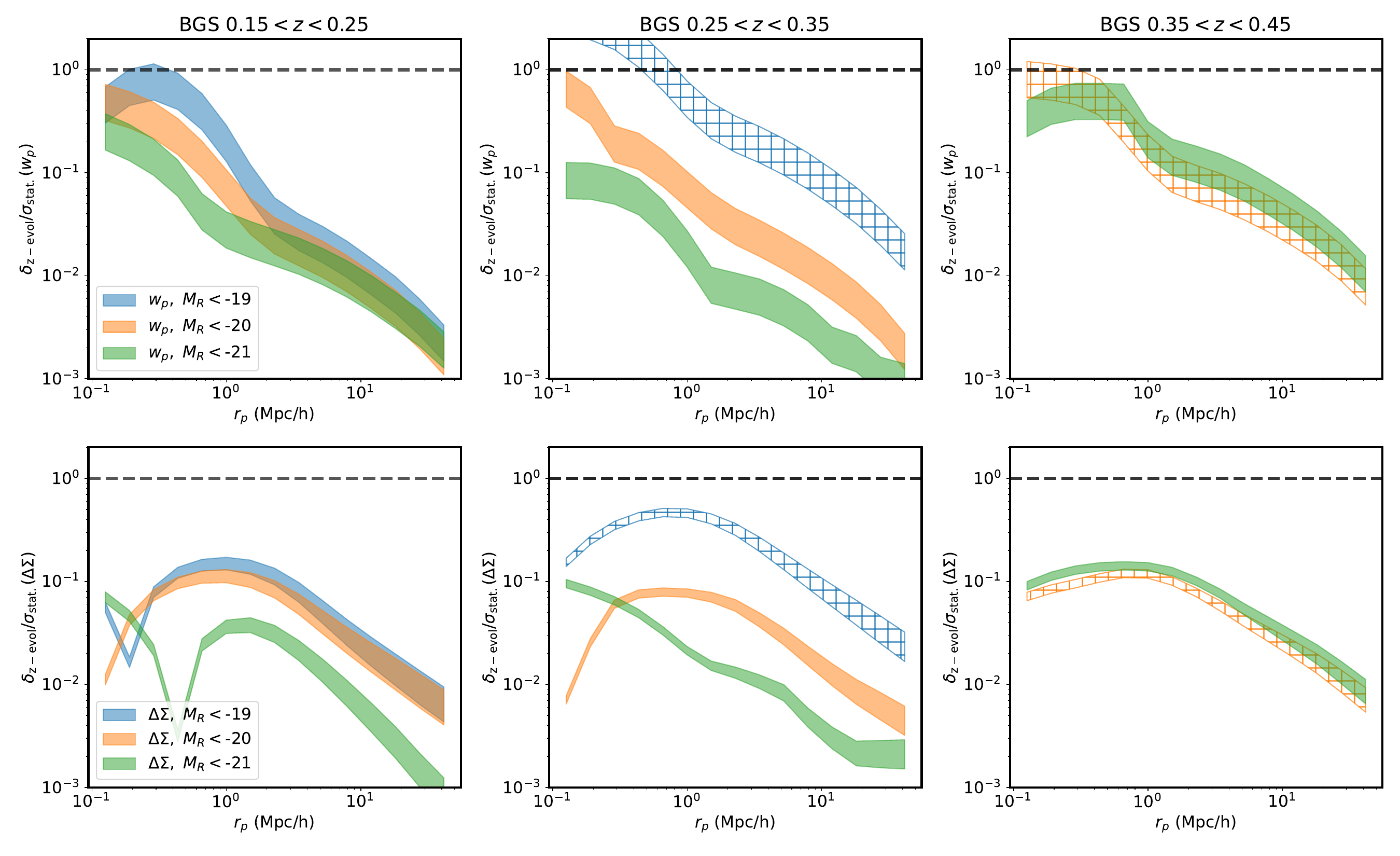}
    \vspace{-0.3cm}
    \caption{The ratio of the redshift evolution effect in the fiducial redshift bins to the expected statistical error for BGS sample with different magnitude cuts. The three columns correspond to the three fiducial BGS lens bins. The top row shows the results for $w_p$, whereas the bottom row shows the results for $\Delta\Sigma$. The three colors show the three different absolute magnitude cuts, while the width of the curves represents the spread between different photometric samples. We have skipped $M_R < -19$ in $0.3 < z < 0.4$ as it is the same sample as $M_R < -20$ at $0.3 < z < 0.4$. We have also used open hatches instead of solid colors for the two bins that are highly incomplete (dashed lines in Figure~\ref{fig:bgs_mrz}). We see that for galaxy lensing, redshift evolution remains insignificant across all scales and magnitude cuts in the fiducial redshift bins. For galaxy clustering, redshift evolution is insignificant at scales greater than $r_p > 1 \, h^{-1}$Mpc. The redshift evolution is less important for brighter samples and tend to be worst for highly incomplete samples.}
    \label{fig:bgs_noise_bias}
\end{figure*}

Figure~\ref{fig:bgs_noise_bias} shows the ratio of $\delta_\mathrm{z-evol}$ to $\sigma_\mathrm{stat.}$ as a function of scale, but we organize the plots differently than for LRGs. The three columns correspond to the three fiducial redshift bins, whereas the the two rows correspond to the two summary statistics. We use three different colors to show three different magnitude cuts. The horizontal dashed lines indicate the threshold where the redshift evolution becomes a statistically significant effect. The widths on the curves again show the spread between different source samples. We have skipped $M_R < -19$ in $0.3 < z < 0.4$ as it is the same sample as $M_R < -20$ at $0.3 < z < 0.4$. We have also used open hatches instead of solid colors for the two bins where the sample is highly incomplete: $M_R < -19$ in $0.2 < z < 0.3$, and $M_R < -20$ in $0.3 < z < 0.4$. These two choices are not relevant for the final cosmology analysis as they do not represent significant gains over brighter magnitude cuts in the same redshift bins. 

We see that the systematics due to redshift evolution are somewhat more significant for BGS than for LRGs. As expected, the significance of redshift evolution decreases with the limiting absolute magnitude. For lensing, redshift evolution remains insignificant across all scales and all magnitude cuts in the fiducial redshift bins. For clustering, redshift evolution is notably more important due to the bias squared scaling. Specifically, redshift evolution can be significant for $w_p$ at the very small scales $r_p < 1 \, h^{-1}$Mpc, but the significance drops as we move to brighter samples and higher completeness. The highly incomplete selections shown in open hatches tend to suffer from worse evolution effects. Thus, for analyses that focus on 2-halo scales and larger, redshift evolution should be insignificant given Y1 statistics. For analyses that aim to leverage the 1-halo scales, more care should be given to evolution effect. 

A new type of analyses using simulation-based models to derive cosmological constraints have recently gained popularity. Some recent examples include for example \cite{2023Lange, 2022bYuan, 2022Zhai, 2021Lange}. While promising to utilize smaller scales than traditional analytic methods, these analyses are limited to redshift snapshots available in the simulation outputs. We assess the redshift evolution systematics for these types of analyses in Appendix~\ref{sec:simbased}.

\section{Discussion and Conclusions}
\label{sec:conclude}

DESI Y1 data are expected to deliver high-precision GGL and clustering measurements and strong constraints on cosmological parameters. It is essential to forecast the statistical errors and characterise the impact of different systematics. 
In this paper, we present the covariance forecasts for DESI Y1 GGL and clustering measurements and assess the significance of lens sample redshift evolution. Our key findings are:
\begin{itemize}
    \item A Gaussian analytical covariance calculation agrees with the galaxy-galaxy lensing error across an ensemble of mock catalogues within around $10\%$, across the majority of scales and redshifts.
    \item Survey footprint effects increase the covariance on the largest scales, especially for the ``cross'' component of galaxy-galaxy lensing. These effects will be important when using B-mode tests on scales comparable to the survey footprint. 
    \item The DESI LRG sample in fiducial bins of $0.4 < z < 0.6$ and $0.6 < z < 0.8$ shows very mild evolution with insignificant effect on the lensing and clustering measurements.
    \item The evolution in the BGS sample is dependent on the absolute magnitude cut, with brighter cuts resulting in less evolution and vice versa. Galaxy lensing should not be sensitive to redshift evolution across all scales, magnitude cuts, and redshift bins in our fiducial setup. Galaxy clustering is more sensitive, where evolution effects can be significant on 1-halo scales at $r_p < 1h^{-1}$Mpc.
\end{itemize}
This paper should be a reference for making analysis choices with DESI Y1 samples, especially in selecting redshift bins and magnitude cuts. For LRGs, we find the fiducial redshift bins to be more than sufficient in controlling the redshift evolution effects. For BGS, we find galaxy--galaxy lensing to be safe given fiducial choices. Galaxy clustering is also safe if the analysis does not utilise the smallest scales. For small-scale analyses that want to maximise sample size, additional redshift bins might be needed. Alternatively, one can adopt different magnitude cuts for different redshift ranges. Specifically, lower redshift bins tend to see less evolution at fixed luminosity and thus can adopt a fainter magnitude cut. Figure~\ref{fig:bgs_noise_bias} acts as a reference plot for making these magnitude choices. 

There are several caveats with this analysis. For simplicity, we do not characterise or report the uncertainties on the HOD fits. This means that the systematic effects we report in Figure~\ref{fig:lrg_noise_bias} and Figure~\ref{fig:bgs_noise_bias} are of their mean expected amplitudes, not their worst-case amplitudes. However, because the best-fit HODs produce continuous and expected behaviors as a function of redshift, we expect the uncertainties of the HOD fits to be small and that our conclusions are robust. Similarly, the HODs themselves are calibrated in finite redshift bins assuming a model at a fixed redshift. We designed these bins to be centered on the simulation output in the BGS analysis to reduce this bias. Nevertheless, a more rigorous examination of redshift evolution is needed beyond Y1. 

Additionally, we have modeled the redshift dependence with an interpolation scheme between a small number of snapshots. While that is sufficient for identifying the overall trends, it potentially introduces spurious evolution effects due to the low number of interpolated points and errors in the best-fit. We suggest a repeat of this analysis using simulated lightcones and properly implemented HOD$(z)$ models. 
Finally, we directly applied the best-fit HODs from \textsc{AbacusSummit} to Dark Emulator, which relies on a different set of simulations and different halo finders. This leads to a small systematic bias that is ignored in this paper.

\section*{Acknowledgements}
This work was supported by grant DE-SC0013718 and under DE-AC02-76SF00515 to SLAC National Accelerator Laboratory, and by the Kavli Institute for Particle Astrophysics and Cosmology. MdlR was supported by a Stanford Science Fellowship at Stanford University. This work was performed in part at the Aspen Center for Physics, which is supported by National Science Foundation grant PHY-2210452.

This material is based upon work supported by the U.S. Department of Energy (DOE), Office of Science, Office of High-Energy Physics, under Contract No. DE–AC02–05CH11231 and DE-SC0010107, and by the National Energy Research Scientific Computing Center, a DOE Office of Science User Facility under the same contract. Additional support for DESI was provided by the U.S. National Science Foundation (NSF), Division of Astronomical Sciences under Contract No. AST-0950945 to the NSF’s National Optical-Infrared Astronomy Research Laboratory; the Science and Technology Facilities Council of the United Kingdom; the Gordon and Betty Moore Foundation; the Heising-Simons Foundation; the French Alternative Energies and Atomic Energy Commission (CEA); the National Council of Science and Technology of Mexico (CONACYT); the Ministry of Science and Innovation of Spain (MICINN), and by the DESI Member Institutions: \url{https://www.desi.lbl.gov/collaborating-institutions}. Any opinions, findings, and conclusions or recommendations expressed in this material are those of the author(s) and do not necessarily reflect the views of the U. S. National Science Foundation, the U. S. Department of Energy, or any of the listed funding agencies.

The authors are honored to be permitted to conduct scientific research on Iolkam Du’ag (Kitt Peak), a mountain with particular significance to the Tohono O’odham Nation.

This work made use of Astropy:\footnote{http://www.astropy.org} a community-developed core Python package and an ecosystem of tools and resources for astronomy \citep{astropy:2013, astropy:2018, astropy:2022}.

\section*{Data Availability}

 The simulation data are available at \url{https://abacussummit.readthedocs.io/en/latest/}. The \ahod\ code package is publicly available as a part of the \textsc{abacusutils} package at \url{https://github.com/abacusorg/abacusutils}. Example usage can be found at \url{https://abacusutils.readthedocs.io/en/latest/hod.html}.
All mock products will be made available at \url{https://data.desi.lbl.gov}.

 All data points shown in the published graph will be also available at \url{https://doi.org/10.5281/zenodo.10724372}.




\bibliographystyle{mnras}
\bibliography{bibi} 

\appendix

\section{Gaussian covariance of projected correlations}
\label{sec:covformulae}

In this section we detail our determination of the analytical covariance for the projected correlation functions, $\Delta\Sigma(r_p)$ and $w_p(r_p)$.

\subsection{Relation of projected correlations to power spectra}

The projected mass density around lens galaxies at projected separation $\vR$ is determined by the 3D galaxy-mass cross-correlation function $\xi_{gm}(\vr)$ as,
\begin{equation}
  \Sigma(\vR) = \overline{\rho}_m \int_{-\infty}^\infty dr_\pi \, \left[ 1 + \xi_{gm}(\vR,r_\pi) \right] ,
\label{eq:projmass}
\end{equation}
where $r_\pi$ is the line-of-sight separation and $\overline{\rho}_m$ is the mean cosmic matter density.  The surface mass density contrast is then found by averaging the projected density over directions in the plane of the sky defined by $\phi$, such that,
\begin{equation}
\begin{split}
  \Delta \Sigma(r_p) &= \overline{\Sigma}(<r_p) - \Sigma(r_p) \\
  &= \frac{2}{r_p^2} \int_0^{r_p} dr_p' \, r_p' \, \Sigma(r_p') - \Sigma(r_p) \\
  &= \overline{\rho}_m \int_{-\infty}^\infty dr_\pi \, \left[ \frac{2}{r_p^2} \int_0^{r_p} dr_p' \, r_p' \, \int_0^{2\pi} \frac{d\phi'}{2\pi} \, \xi_{gm}(\vR',r_\pi) \right. \\
  &\hspace{5mm}- \left. \int_0^{2\pi} \frac{d\phi}{2\pi} \, \xi_{gm}(\vR,r_\pi) \right] ,
\end{split}
\end{equation}
after substituting in Eq.\ref{eq:projmass}.  We can express this in terms of the galaxy-mass cross-power spectrum by using $\xi_{gm}(\vr) = \intk \, P_{gm}(\vk) \, e^{-i\vk \cdot \vr}$, obtaining,
\begin{equation}
\begin{split}
   \Delta \Sigma(r_p) &= \overline{\rho}_m \intk \, P_{gm}(\vk) \left[ \int_{-\infty}^\infty dr_\pi \, e^{-i k_\parallel r_\pi} \right] \, \times \\
   &\hspace{5mm} \left[ \frac{2}{r_p^2} \int_0^{r_p} dr_p' \, r_p' \, \int_0^{2\pi} \frac{d\phi'}{2\pi} \, e^{-i\vk_\perp \cdot \vR'} - \int_0^{2\pi} \frac{d\phi}{2\pi} \, e^{-i \vk_\perp \cdot \vR} \right] \\
   &= \overline{\rho}_m \intk \, P_{gm}(\vk) \, \times \\
   &\hspace{5mm} \left[ L_\parallel \, \tilde{\delta}_D(k_\parallel) \right] \left[ \frac{2}{r_p^2} \int_0^{r_p} dr_p' \, r_p' \, J_0(k_\perp r_p') - J_0(k_\perp r_p) \right] \\
   &= \overline{\rho}_m \intvkperp \, P_{gm}(\vk_\perp) \, J_2(k_\perp r_p) ,
\label{eq:dsig}
\end{split}
\end{equation}
where $\vk_\perp$ is the 2D Fourier wave vector in the plane of the sky, $k_\parallel$ is the wave number in the line-of-sight direction, $L_\parallel$ is the thickness of the redshift slice, and $J_n$ indicates a Bessel function of the first kind.  The projected galaxy correlation function at projected separation $\vR$ is defined by,
\begin{equation}
  w_p(\vR) = \int_{-r_{\pi,{\rm max}}}^{r_{\pi,{\rm max}}} dr_\pi \, \xi_{gg} (\vR,r_\pi) ,
\end{equation}
where $r_{\pi,{\rm max}}$ is the upper limit used in the measurement, and $\xi_{gg}(\vr)$ is the 3D galaxy correlation function.  We can express this in terms of the galaxy auto-power spectrum by using $\xi_{gg}(\vr) = \intk \, P_{gg}(\vk) \, e^{-i\vk \cdot \vr}$, obtaining,
\begin{equation}
  w_p(\vR) = \intk P_{gg}(\vk) \, e^{-i\vk_\perp \cdot \vR} \int_{-r_{\pi, {\rm max}}}^{r_{\pi,{\rm max}}} dr_\pi\, e^{-i k_\parallel r_\pi} .
\end{equation}
Averaging the measurement over directions $\phi$ we find,
\begin{equation}
  w_p(r_p) = 2 r_{\pi,{\rm max}} \intk \, P_{gg}(\vk) \, J_0(k_\perp r_p) \, j_0(k_\parallel r_{\pi,{\rm max}}) .
\label{eq:wppk}
\end{equation}
where $j_0$ is a spherical Bessel function.  In the Limber approximation we suppose that only tangential modes contribute to the projected clustering, which allows us to simplify Eq.\ref{eq:wppk} to the form,
\begin{equation}
\begin{split}
  w_p^{\rm Lim}(r_p) &= \intvkperp P_{gg}(\vk_\perp) \, J_0(k_\perp r_p) \times 2 r_{\pi,{\rm max}} \, \intkpar \, j_0(k_\parallel r_{\pi, {\rm max}}) \\
&= \intvkperp \, P_{gg}(\vk_\perp) \, J_0(k_\perp r_p) .
\end{split}
\label{eq:wplimpk}
\end{equation}

\subsection{Covariance of projected correlations}

We may deduce the covariance of $\Delta\Sigma(r_p)$ using Eq.\ref{eq:dsig} \citep[for similar treatments, see][]{2017MNRAS.471.3827S, 2018MNRAS.478.4277S, 2018MNRAS.479.1240D, 2020A&A...642A.158B},
\begin{equation}
\begin{split}
  & {\rm Cov} \left[ \Delta\Sigma(r_p) , \Delta\Sigma(r_p') \right] \\
  &= \overline{\rho}_m^2 \intvkperp \intvkpperp \, {\rm Cov} \left[ P_{gm}(\vk_\perp) , P_{gm}(\vk_\perp') \right] \, J_2(k_\perp r_p) \, J_2(k_\perp' r_p') \\
  &= \frac{\overline{\rho}_m^2}{A_s} \intvkperp \, {\rm Cov} \left[ P_{gm}(\vk_\perp) , P_{gm}(\vk_\perp) \right] \, J_2(k_\perp r_p) \, J_2(k_\perp r_p') ,
\end{split}
\end{equation}
where $A_s = \Omega_s \, \chi_{\rm eff}^2$ is the projected survey area at the effective lens distance $\chi_{\rm eff}$.

We now use the relations between the spatial and angular power spectra for a narrow redshift slice,
\begin{equation}
  C^{\rm Nar}_{g\kappa}(\ell) = \overline{\rho}_m \, \frac{\overline{\Sigma_c^{-1}}(\chi_{\rm eff})}{\chi_{\rm eff}^2} \, P_{gm} \left( \frac{\ell}{\chi_{\rm eff}} \right) ,
\label{eq:clgknar}
\end{equation}
where $\ell$ is the angular wave number and the general expression for the covariances between the angular power spectra of different fields (labelled $A,B,C,D$) and samples (labelled $i,j,k,l$) \citep{2017MNRAS.470.2100K},
\begin{equation}
\begin{split}
  {\rm Cov} \left[ C_{AB}^{ij} , C_{CD}^{kl} \right] &= \left[ C_{AC}^{ik} + \delta^K_{ik} \, \delta^K_{AC} \, N_A^i \right] \left[ C_{BD}^{jl} + \delta^K_{jl} \, \delta^K_{BD} \,  N_B^j \right] \\
  &+ \left[ C_{AD}^{il} + \delta^K_{il} \,  \delta^K_{AD} \, N_A^i \right] \left[ C_{BC}^{jk} + \delta^K_{jk} \, \delta^K_{BC} \, N_B^j \right]
\label{eq:covgeneral}
\end{split}
\end{equation}
where $N$ indicates the noise auto-power spectrum of a field.  Applying Eq.\ref{eq:clgknar} and Eq.\ref{eq:covgeneral},
\begin{equation}
\begin{split}
    &{\rm Cov} \left[ P_{gm}(\vk_\perp) , P_{gm}(\vk_\perp) \right] \\
    &= \frac{\chi_{\rm eff}^4}{\overline{\rho}_m^2 \, \left[ \overline{\Sigma_c^{-1}}(\chi_{\rm eff}) \right]^2} \, {\rm Cov} \left[ C^{\rm Nar}_{g\kappa}(\vk_\perp \chi_{\rm eff}) , C^{\rm Nar}_{g\kappa}(\vk_\perp \chi_{\rm eff}) \right] \\
    &= P^2_{gm}(k_\perp) + \frac{\chi_{\rm eff}^4}{\overline{\rho}_m^2 \, \left[ \overline{\Sigma_c^{-1}}(\chi_{\rm eff}) \right]^2} \left[ C_{\kappa\kappa}(k_\perp \chi_{\rm eff}) + N_\kappa \right] \left[ C^{\rm Nar}_{gg}(k_\perp \chi_{\rm eff}) + N_g \right] .
\end{split}
\end{equation}
The expression for the covariance is then,
\begin{equation}
\begin{split}
  & {\rm Cov} \left[ \Delta\Sigma(r_p) , \Delta\Sigma(r_p') \right] = \frac{1}{A_s} \intkperp J_2(k_\perp r_p) J_2(k_\perp r_p') \, \times \\
  &\left\{ \overline{\rho}_m^2 P^2_{gm}(k_\perp) + \frac{\chi_{\rm eff}^4}{\left[ \overline{\Sigma_c^{-1}}(\chi_{\rm eff}) \right]^2} \left[ C_{\kappa\kappa}(k_\perp \chi_{\rm eff}) + N_\kappa \right] \left[ C^{\rm Nar}_{gg}(k_\perp \chi_{\rm eff}) + N_g \right] \right\}
\end{split}
\end{equation}
Similarly for the ``cross'' component,
\begin{equation}
\begin{split}
  &{\rm Cov} \left[ \Delta\Sigma_\times(r_p) , \Delta\Sigma_\times(r_p') \right] = \frac{1}{A_s} \, \frac{\chi_{\rm eff}^4}{\left[ \overline{\Sigma_c^{-1}}(\chi_{\rm eff}) \right]^2} \, \times \\
  &\hspace{5mm} \intkperp \, J_2(k_\perp r_p) \, J_2(k_\perp r_p') \, N_\kappa \left[ C^{\rm Nar}_{gg}(k_\perp \chi_{\rm eff}) + N_g \right] .
\label{eq:covdsigcross}
\end{split}
\end{equation}
The covariance of $w_p(r_p)$ follows from Eq.\ref{eq:wppk},
\begin{equation}
\begin{split}
  &{\rm Cov} \left[ w_p(r_p), w_p(r_p') \right] = 4 r_{\pi,{\rm max}}^2 \intk \intkp \\
  &\hspace{5mm} {\rm Cov} \left[ P_{gg}(\vk), P_{gg}(\vk') \right] J_0(k_\perp r_p) \, J_0(k'_\perp r_p') \, j_0(k_\parallel r_{\pi,{\rm max}}) \, j_0(k'_\parallel r_{\pi,{\rm max}}) .
\end{split}
\end{equation}
Using ${\rm Cov} \left[ P_{gg}(\vk), P_{gg}(\vk') \right] = 2 \, \left[ P_{gg}(\vk) + \frac{1}{n_g} \right]^2 \, \tilde{\delta}_D(\vk-\vk')$, where the factor of 2 accounts for the fact that the modes $-\vk$ and $\vk$ are not independent, this becomes,
\begin{equation}
\begin{split}
  {\rm Cov} \left[ w_p(r_p), w_p(r_p') \right] &= \frac{8 r_{\pi,{\rm max}}^2}{V_s} \intk \left[ P_{gg}(\vk) + \frac{1}{n_g} \right]^2 \times \\
  &\, J_0(k_\perp r_p) \, J_0(k_\perp r_p') \, j_0^2(k_\parallel r_{\pi,{\rm max}}) ,
\end{split}
\end{equation}
where $V_s$ is the survey volume.  In the Limber approximation where only tangential modes contribute to the covariance,
\begin{equation}
\begin{split}
  &{\rm Cov}^{\rm Lim} \left[ w_p(r_p), w_p(r_p') \right] \\
  &= \frac{8 r_{\pi,{\rm max}}^2}{V_s} \intvkperp \left[ P_{gg}(k_\perp) + \frac{1}{n_g} \right]^2 \, J_0(k_\perp r_p) \, J_0(k_\perp r_p') \, \times \\
  &\hspace{5mm} \intkpar j_0^2(k_\parallel r_{\pi,{\rm max}}) \\
  &= \frac{4 r_{\pi,{\rm max}}}{V_s} \intkperp \left[ P_{gg}(k_\perp) + \frac{1}{n_g} \right]^2 \, J_0(k_\perp r_p) \, J_0(k_\perp r_p') \\
  &= \frac{4 r_{\pi,{\rm max}}}{V_s} \intkperp \left[ P_{gg}^2(k_\perp) + \frac{2 P_{gg}(k_\perp)}{n_g} \right] \, J_0(k_\perp R) \, J_0(k_\perp r_p') \\
  &\hspace{5mm} + \frac{2 r_{\pi,{\rm max}}}{\pi r_p \, n_g^2 \, V_s} \, \delta_D(r_p-r_p') .
\label{eq:wpcov}
\end{split}
\end{equation}
Eq.\ref{eq:wpcov} is identical to Eq.31 in \cite{2015MNRAS.451.1418M}, neglecting the term ``$w^{\rm Lim}_{gg}(r_p)$'' in their expression.  This term arises if we use a Poisson rather than a Gaussian model for the galaxy statistics, and can be neglected on large scales.

\subsection{Cross-covariances involving projected correlations}

The cross-covariance between $\Delta\Sigma(r_p)$ and $w_p(r_p)$ follows from Eq.\ref{eq:dsig} and Eq.\ref{eq:wplimpk},
\begin{equation}
\begin{split}
  &{\rm Cov} \left[ \Delta\Sigma(r_p), w_p^{\rm Lim}(r_p') \right] = \\
  &\hspace{5mm} \frac{\overline{\rho}_m}{A_s} \, \intvkperp \, {\rm Cov} \left[ P_{gm}(\vk_\perp) , P_{gg}(\vk_\perp) \right] \, J_2(k_\perp r_p) \, J_0(k_\perp r_p') .
\end{split}
\end{equation}
We now use the relation for narrow redshift slice,
\begin{equation}
  C^{\rm Nar}_{gg}(\ell) = \frac{1}{\chi_{\rm eff}^2 \, L_\parallel} P_{gg}\left(\frac{\ell}{\chi_{\rm eff}}\right) .
\label{eq:clggnar}
\end{equation}
Applying Eq.\ref{eq:clggnar}, Eq.\ref{eq:clgknar} and Eq.\ref{eq:covgeneral},
\begin{equation}
\begin{split}
    &{\rm Cov} \left[ P_{gm}(\vk_\perp) , P_{gg}(\vk_\perp) \right] \\
    &= \frac{\chi_{\rm eff}^4 \, L_\parallel}{\overline{\rho}_m \, \overline{\Sigma_c^{-1}}(\chi_{\rm eff})} \, {\rm Cov} \left[ C^{\rm Nar}_{g\kappa}(k_\perp \chi_{\rm eff}) , C^{\rm Nar}_{gg}(k_\perp \chi_{\rm eff}) \right] \\
    &= \frac{\chi_{\rm eff}^4 \, L_\parallel}{\overline{\rho}_m \, \overline{\Sigma_c^{-1}}(\chi_{\rm eff})} \times 2 \, \left[ C^{\rm Nar}_{gg}(k_\perp \chi_{\rm eff}) + N_g \right] C^{\rm Nar}_{g\kappa}(k_\perp \chi_{\rm eff}) \\
    &= 2 \left[ P_{gg}(k_\perp) + \frac{1}{n_g} \right] P_{gm}(k_\perp) .
\end{split}
\end{equation}
Hence, the cross-covariance expression is,
\begin{equation}
\begin{split}
  &{\rm Cov} \left[ \Delta\Sigma(r_p), w_p^{\rm Lim}(r_p') \right] = \\
  &\frac{2 \, \overline{\rho}_m}{A_s} \, \intkperp \, \left[ P_{gg}(k_\perp) + \frac{1}{n_g} \right] P_{gm}(k_\perp) \, J_2(k_\perp r_p) \, J_0(k_\perp r_p') .
\end{split}
\end{equation}

\section{Testing footprint effects on the covariance using lognormal mocks}
\label{sec:covfootprint}

In this section we use lognormal mock catalogues to calibrate the impact of the survey footprint on the covariance of galaxy-galaxy lensing statistics.  For some previous investigations of this issue, we refer the reader to \cite{2004A&A...413..465K, 2011ApJ...734...76S, 2019MNRAS.486...52S, 2021MNRAS.508.3125F, 2021A&A...646A.129J}.  Footprint effects cannot be readily included in analytical treatments of the covariance, even for the Gaussian term.  However, numerical investigations allow the significance of the effects to be explored.

Lognormal realisations are a useful resource for such investigations because a large ensemble of simulations can be cheaply generated (we use 1000 realisations for our analysis), permitting the differences between the covariances for different configurations to be accurately quantified.  We generated these mocks using the Full-sky Lognormal Astro-fields Simulation Kit \citep[FLASK,][]{2016MNRAS.459.3693X}, adopting the same fiducial cosmology as the Buzzard N-body simulations and a lens redshift distribution and bias evolution matching the DESI configuration.  We simulated a weak lensing source distribution with a density of $10$ arcmin$^{-2}$, sampled from a generic source redshift distribution for $z < 1.85$ which we sub-sampled into the tomographic bins of the three weak lensing surveys.  (We could match the complete source density of KiDS and DES, and we sub-sampled the HSC source density by a factor of 2.)  We also sampled the mocks to match the angular completeness of each of the DESI and weak lensing survey overlap regions illustrated in Fig.\ref{fig:desiy1_surveys_completeness}.

We measured the tangential and cross-components of the galaxy-galaxy lensing signal for each realisation, $\gamma_t$ and $\gamma_\times$, and compared the result to the analytical covariance determination.  For the purposes of this comparison, the analytical covariances were computed using the effective convergence and galaxy model power spectra output by the FLASK code, at the corresponding angular resolution, which capture the lognormal effects.  These power spectra differ from the models used to determine the analytical covariance of the Buzzard simulations, which is why the numerical FLASK covariance cannot be used directly as the data covariance.

\begin{figure}
\includegraphics[width=0.45\textwidth]{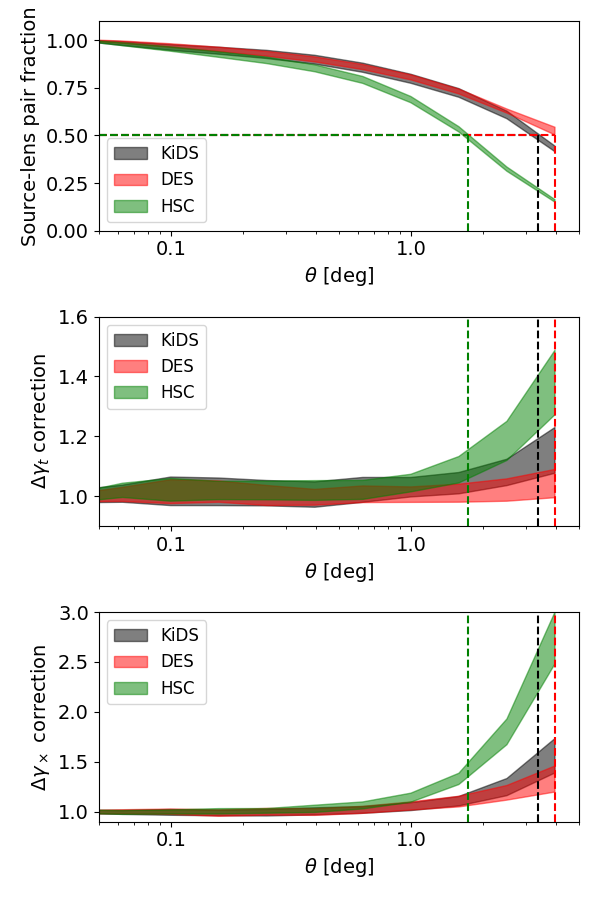}
\caption{The impact of survey geometry on the error in the tangential and cross-components of the average shear ($\gamma_t$ and $\gamma_\times$).  The upper panel shows the fraction of observed source-lens pairs as a function of angular separation $\theta$, compared to the theoretical pair count that would be observed in the absence of survey boundary effects.  Results are shown for KiDS, DES and HSC DESI overlap footprints, where the band indicates the standard deviation across different pairs of lens and source samples.  The angular scale corresponding to losing $50\%$ of pairs is indicated by the vertical dashed line.  The middle and lower panel display the ratio of the standard deviation of the average shear measurements across the realisations, to the analytical covariance error.  The bands again indicate the standard deviation of results across the different lens and source samples.}
\label{fig:pairfrac_witherrors}
\end{figure}

The results of this galaxy-galaxy lensing error comparison are displayed in Fig.\ref{fig:pairfrac_witherrors}, for the overlap footprints between DESI and KiDS, DES and HSC.  The upper panel shows the observed source-lens pairs as a function of angular separation, as a fraction of the theoretical pair count that would be observed in the absence of survey boundary effects.  The angular scale where $50\%$ of pairs are lost is indicated by the vertical dashed line, which occurs at $3.39^\circ$, $4.29^\circ$ and $1.72^\circ$ for the KiDS, DES and HSC overlaps, respectively.  These relative scales agree well with the survey geometries displayed in Fig.\ref{fig:desiy1_surveys_completeness}.  The remaining two panels of Fig.\ref{fig:pairfrac_witherrors} display the ratio of the standard deviation of the $\gamma_t$ and $\gamma_\times$ measurements across the realisations, to the prediction of the analytical covariance.  The width of the bands indicate the variation across the different combinations of lens and source tomographic bins, which are combined in this plot for clarity.  The two covariance determinations agree well on small scales, but we find that the analytical calculation underestimates the survey covariance on large scales, with this effect being most significant for $\gamma_\times$.  We can interpret this finding by considering the ``mixing'' of E-mode and B-mode convergence power in the contributions to the Gaussian covariance including a survey footprint: the contribution of the (larger) E-mode power to $\gamma_\times$ exceeds that of the (smaller) B-mode power to $\gamma_t$.  We find that at the angular scales noted above where $50\%$ of pairs are lost, the $\gamma_t$ error increases by around $10\%$ or less, and the $\gamma_\times$ error increases by around $30\%$.

In our calculation of the analytical covariances for these survey footprints, we include these effects by multiplying the diagonal errors by this correction factor, whilst maintaining the same cross-correlation between the off-diagonal elements as predicted by the analytical calculation.  (The corrections for the off-diagonal covariances are noisy, as the denominators are small.)  For the $\Delta\Sigma$ covariances, we cannot compute the correction directly as a function of $R$, since the FLASK simulations are purely angular.  Hence in this case we interpolate the correction at the corresponding angular separation of each lens redshift bin, $\theta = R/\chi(z_l)$.

\begin{figure}
\includegraphics[width=0.45\textwidth]{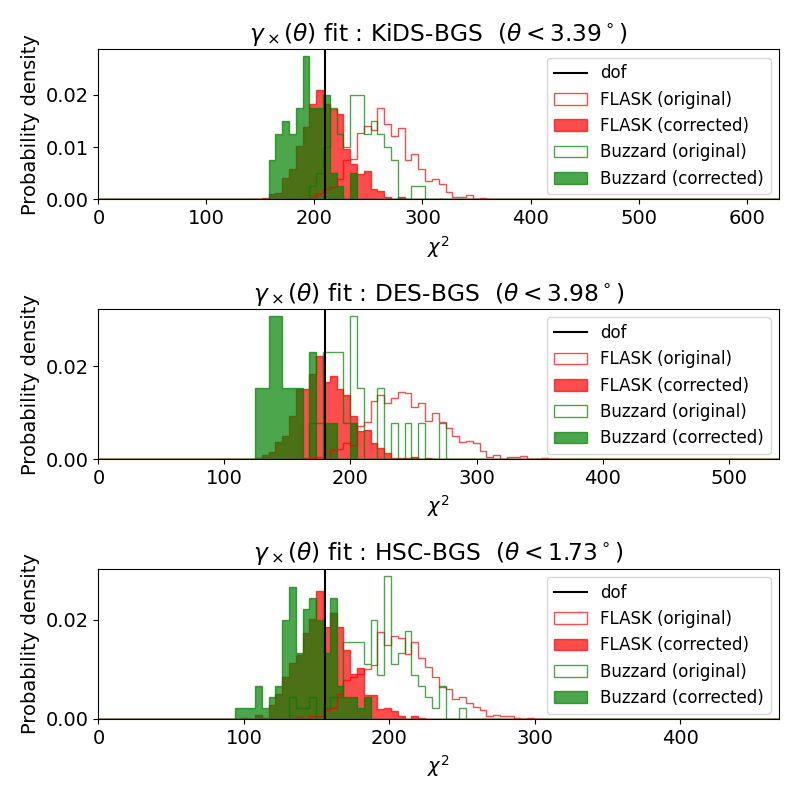}
\caption{The distribution of $\chi^2$ values obtained when performing a ``B-mode test'' (that $\gamma_\times = 0$) using the Buzzard and FLASK simulations including the KiDS, DES and HSC overlap footprints.  The open and solid histograms correspond respectively to using the original analytical covariance, and the covariance corrected for survey footprint effects using the FLASK lognormal mocks.  The vertical line corresponds to the number of degrees of freedom in each case.}
\label{fig:chisq_gx_flask}
\end{figure}

We tested the efficacy of these corrections by applying them to a ``B-mode test'' of the Buzzard and FLASK mocks.  This test verifies that the $\gamma_\times$ measurements are consistent with zero, using the assumed covariance combining all source and lens bins.  We consider analyses adopting both the original analytical covariance, and the corrected covariance described above.  The distribution of $\chi^2$ values across the different realisations, corresponding to these cases, is shown in Fig.\ref{fig:chisq_gx_flask}. We assume the maximum fitting scales indicated in Fig.\ref{fig:pairfrac_witherrors}, at which $50\%$ of source-lens pairs are excluded by the survey boundary.  We find that if the original (uncorrected) covariance is used, the $\chi^2$ values are over-estimated compared to the number of degrees of freedom, owing to the under-estimate of the covariance, such that the $\gamma_\times = 0$ hypothesis is erroneously rejected for many realisations.  However, following the correction to the covariance based on the FLASK lognormal mocks, an appropriate $\chi^2$ distribution is recovered again.

\section{Simulation-based analyses}
\label{sec:simbased}
In this section, we characterize the significance of redshift evolution effects for simulation-based analyses. For DESI, two simulation suites are considered for such analyses: \textsc{AbacusSummit} \citep{2021Maksimova} and \textsc{Aemulus} \citep{2019DeRose}. The available relevant redshift snapshots are displayed at the bottom of Figure~\ref{fig:lens_nz} in short vertical lines. For the BGS sample, the \textsc{AbacusSummit} snapshots are unfortunately lying at the edge of the fiducial redshift bins. Using these snapshots for the fiducial bins is expected to result in large systematic biases. Thus, it is necessary for simulation-based analyses to use custom non-fiducial redshift bins for the BGS sample. 

\begin{figure*}
    \hspace{-0.7cm}
    \includegraphics[width=1.03\textwidth]{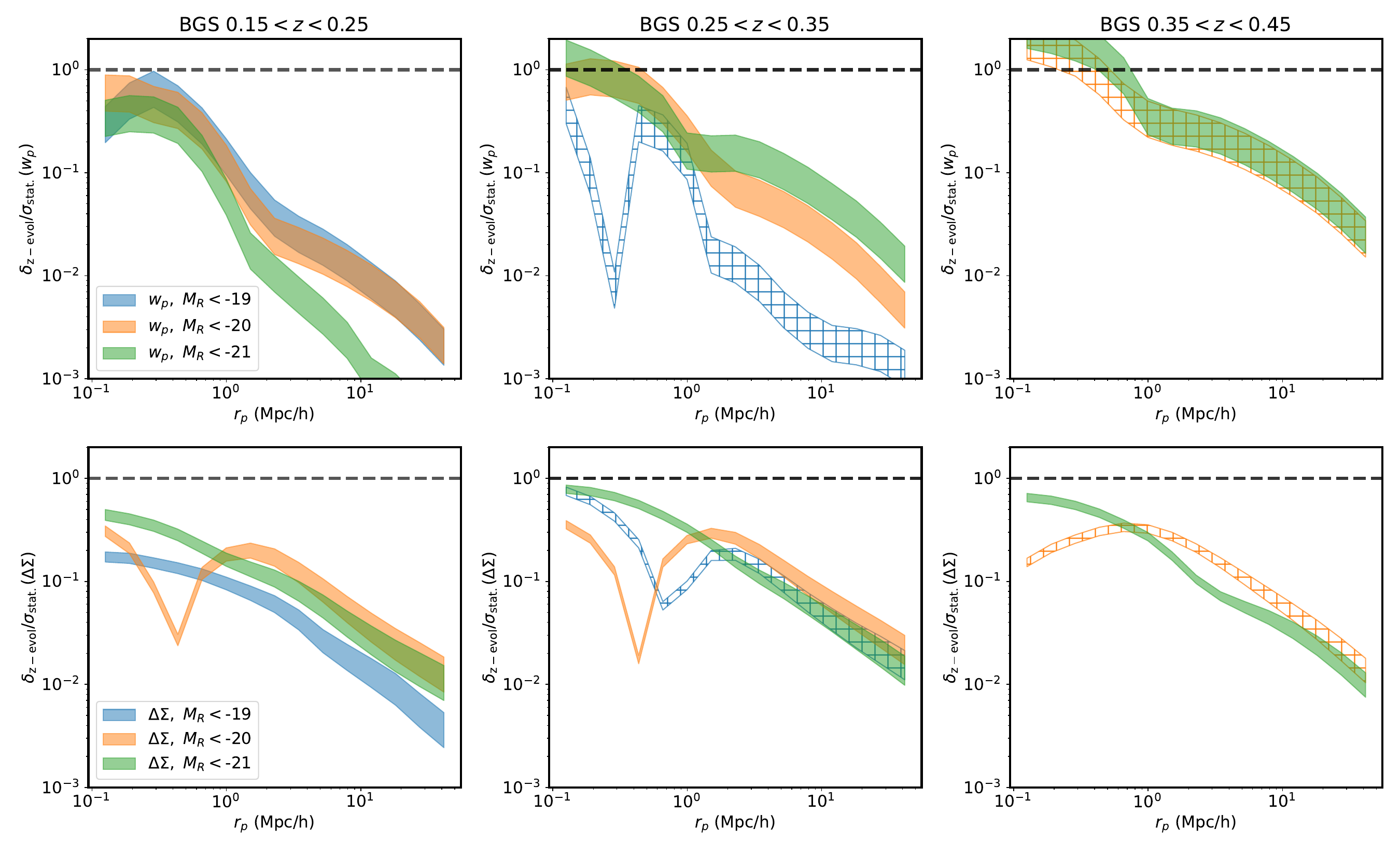}
    \vspace{-0.3cm}
    \caption{The ratio of the redshift evolution effect in the offset redshift bins to the expected statistical error for BGS sample in a simulation-based analysis. The three columns correspond to the three new BGS lens bins centered on \textsc{AbacusSumit} snapshots ($z = 0.2, 0.3, 0.4$). The top row shows the results for $w_p$, whereas the bottom row shows the results for $\Delta\Sigma$. The three colors show the three different absolute magnitude cuts, while the width of the curves represents the spread between different photometric samples. The color scheme is identical to Figure~\ref{fig:bgs_noise_bias}. We see that redshift evolution is insignificant at scales greater than $r_p > 1 \, h^{-1}$Mpc.} 
    \label{fig:bgs_noise_bias_sim}
\end{figure*}

Given that the relevant \textsc{AbacusSummit} snapshots are at $z = 0.2, 0.3, 0.4$, the simplest redshift binning scheme for \textsc{AbacusSummit}-based analyses is $0.15 < z < 0.25$, $0.25 < z < 0.35$, and $0.35 < z < 0.45$. 
Figure~\ref{fig:bgs_noise_bias_sim} shows the ratio between the redshift evolution bias and the cumulative noise-to-signal of the BGS sample in this alternative binning scheme and where the theory predictions are made at $z = 0.2, 0.3, 0.4$. We have noted the custom offset bins in the title of the panels. We have again highlighted the incomplete selections with open hatches. 

Comparing to Figure~\ref{fig:bgs_noise_bias}, we see larger evolution effects due to the fact that we are limited to fixed redshifts instead of the mean sample redshift. However, the qualitative trends in terms of scale and magnitude cuts remain the same, as we expect. 

For galaxy--galaxy lensing, we see that the bias due to redshift evolution continues to be insignificant relative to statistical noise. For galaxy clustering, the evolution effects become important at scales $r_p \sim 1 \, h^{-1}$Mpc and below, especially at higher redshift. This suggests that simulation-based analyses that intend to utilize $r_p < 1 \, h^{-1}$Mpc should either carefully account for evolution systematics at the smallest scales, or use narrower redshift bins to reduce the effect of redshift evolution, or change the effective redshift.

\section{Author Affiliations}
\label{sec:affiliations}
$^{1}$Kavli Institute for Particle Astrophysics and Cosmology, Stanford University, 452 Lomita Mall, Stanford, CA 94305, USA\\
$^{2}$SLAC National Accelerator Laboratory, 2575 Sand Hill Road, Menlo Park, CA  94025, USA\\
$^{3}$Centre for Astrophysics and Supercomputing, Swinburne University of Technology, Hawthorn, VIC 3122, Australia\\
$^4$Perimeter Institute for Theoretical Physics, 31 Caroline St. North, Waterloo, ON N2L 2Y5, Canada\\
$^5$Waterloo Centre for Astrophysics, University of Waterloo, 200 University Ave W, Waterloo, ON N2L 3G1, Canada\\
$^6$Department of Physics, University of Michigan, Ann Arbor, MI 48109, USA\\
$^7$The Ohio State University, Columbus, 43210 OH, USA\\
$^8$Department of Astronomy and Astrophysics, University of California, Santa Cruz, 1156 High Street, Santa Cruz, CA 95065, USA\\
$^{9}$Physics Division, Lawrence Berkeley National Laboratory, Berkeley, CA 94720, USA\\
$^{10}$Physics Dept., Boston University, 590 Commonwealth Avenue, Boston, MA 02215, USA\\
$^{11}$Argonne National Laboratory, High-Energy Physics Division, 9700 S. Cass Avenue, Argonne, IL 60439, USA\\
$^{12}$Department of Physics \& Astronomy, University College London, Gower Street, London, WC1E 6BT, UK\\
$^{13}$Departamento de F\'{\i}sica, Centro de Investigaci\'{o}n y de Estudios Avanzados del IPN, Av. Polit\'{e}cnico 2508, Col. Sn. Pedro Zacatenco, Del. Gustavo A. Madero, 07010 CDMX, M\'{e}xico\\
$^{14}$Department of Physics, The University of Texas at Dallas, Richardson, TX 75080, USA\\
$^{15}$Institut d'Estudis Espacials de Catalunya (IEEC), 08034 Barcelona, Spain\\
$^{16}$Institute of Cosmology $\&$ Gravitation, University of Portsmouth, Dennis Sciama Building, Portsmouth, PO1 3FX, UK\\
$^{17}$Aix Marseille Univ, CNRS/IN2P3, CPPM, Marseille, France\\
$^{18}$Departament de F\'{i}sica, Serra H\'{u}nter, Universitat Aut\`{o}noma de Barcelona, 08193 Bellaterra (Barcelona), Spain\\
$^{19}$NSF's NOIRLab, 950 N. Cherry Ave., Tucson, AZ 85719, USA\\
$^{20}$National Astronomical Observatories, Chinese Academy of Sciences, A20 Datun Rd., Chaoyang District, Beijing, 100012, P.R. China\\
$^{21}$Ruhr University Bochum, Faculty of Physics and Astronomy, Astronomical Institute (AIRUB), German Centre for Cosmological Lensing, 44780 Bochum, Germany\\
$^{22}$Department of Physics, Kansas State University, 116 Cardwell Hall, Manhattan, KS 66506, USA\\
$^{23}$Department of Physics and Astronomy, Sejong University, Seoul, 143-747, Korea\\
$^{24}$CIEMAT, Avenida Complutense 40, E-28040 Madrid, Spain\\
$^{25}$Max Planck Institute for extraterrestrial Physics, 85748 Garching, Bayern, Deutschland\\
$^{26}$Department of Physics \& Astronomy, Ohio University, Athens, OH 45701, USA\\
$^{27}$Instituto de F\'{\i}sica, Universidad Nacional Aut\'{o}noma de M\'{e}xico, Cd. de M\'{e}xico C.P. 04510, M\'{e}xico\\
$^{28}$Observatorio Astron\'omico, Universidad de los Andes, Cra. 1 No. 18A-10, Edificio H, CP 111711 Bogot\'a, Colombia\\

\bsp

\label{lastpage}
\end{document}